\documentclass[12pt,preprint]{aastex}

\newcommand{\etal}{{et al.}}

\newcommand{\kms}{$\, {\rm km\,s^{-1}}$}
\newcommand{\lsun}{$L_{\odot}$}
\newcommand{\msun}{\,$M_{\odot\,}$}
\newcommand{\sch}{Schwarzschild\,\,}
\newcommand{\ml}{$\Upsilon$}
\newcommand{\grad}{^{\circ}}
\shorttitle{Schwarzschild models of discrete data}
\shortauthors{Kleyna et al.}

\begin{document}


\title{Constraining the Mass Profiles of Stellar Systems: \sch Modeling of
Discrete Velocity Datasets}


\author{Julio Chanam\'e\altaffilmark{1}, Jan Kleyna\altaffilmark{2}, \& Roeland van der Marel\altaffilmark{1}}
\altaffiltext{1}{Space Telescope Science Institute, 3700 San Martin Dr., Baltimore, MD 21218}
\altaffiltext{2}{Institute for Astronomy, University of Hawaii, 2680 Woodlawn Drive, Honolulu, HI 96822}


%\altaffiltext{1}{Visiting Astronomer, Cerro Tololo Inter-American Observatory.
%CTIO is operated by AURA, Inc.\ under contract to the National Science
%\altaffiltext{2}{Society of Fellows, Harvard University.}
%\altaffiltext{3}{present address: Center for Astrophysics,
%    60 Garden Street, Cambridge, MA 02138}
%\altaffiltext{4}{Visiting Programmer, Space Telescope Science Institute}
%\altaffiltext{5}{Patron, Alonso's Bar and Grill}


\begin{abstract}

We present a new \sch orbit-superposition code that is designed to
model discrete datasets composed of velocity measurements of
individual kinematic tracers in a dynamical system. This constitutes
an extension of previous implementations that can only address
continuous data in the form of (the moments of) velocity
distributions, thus avoiding potentially important losses of
information due to data binning. Furthermore, the code can handle any
combination of available velocity components, i.e., only line-of-sight
velocities, only proper motions, or a combination of both. It can also
handle a combination of discrete and continuous data. The code
determines the combination of orbital mass weights (representing the
distribution function) as a function of the three integrals of motion
$E,L_z,$ and $I_3$ that best reproduces, in a maximum-likelihood
sense, the available kinematic and photometric observations in a given
axisymmetric gravitational potential. The overall best fit is the one
that maximizes the likelihood over a parameterized set of trial
potentials. The fully numerical approach ensures considerable freedom
on the form of the distribution function $f(E,L_z,I_3)$. This allows a
very general modeling of the orbital structure, thus avoiding
restrictive assumptions about the degree of (an)isotropy of the
orbits. We describe the implementation of the discrete code and
present a series of tests of its performance based on the modeling of
simulated (i.e., artificial) datasets generated from a known
distribution function. We explore pseudo-datasets with varying degrees
of overall rotation and different inclinations on the plane of the
sky, and study the results as a function of relevant observational
variables such as the size of the dataset and the type of velocity
information available. We find that the discrete \sch code recovers
the original orbital structure, mass-to-light ratio, and inclination
of the input datasets to satisfactory accuracy, as quantified by
various statistics. The code will be valuable, e.g., for modeling
stellar motions in Galactic globular clusters, and modeling the
motions of individual stars, planetary nebulae, or globular clusters
in nearby galaxies. This can shed new light on the total mass
distributions of these systems, with central black holes and dark
matter halos being of particular interest.

\end{abstract}





\section{Introduction}
\label{sec.intro}

The study of the internal dynamics of stellar systems plays an
essential role in astronomy.  From the observed positions and
velocities of the stars in galaxies and globular clusters it is
possible to infer their total (dark+luminous) mass distribution,
which, in particular, provides information on the presence and
properties of dark halos and massive black holes. In turn, this
structural knowledge constrains theories for the formation and
evolution of these systems.

The dynamical state of a stellar system is determined by its phase
space distribution function, $f({\vec r}, {\vec v})$, which counts the
stars as a function of position ${\vec r}$ and velocity ${\vec v}$.
Typically, however, only three of the six phase-space coordinates are
available observationally: the projected sky position $(x',y')$, and
the velocity $v_{z'}$ along the line of sight (LOS). Proper motion
observations can provide the additional velocities $(v_{x'},v_{y'})$,
but such data are generally not available (with the notable exception
of some Galactic globular clusters). To make progress with the limited
information available, the dynamical theorist is often forced to make
simplifying assumptions about geometry (e.g., that the system is
spherical) or about the velocity distribution (e.g., that it is
isotropic). Such assumptions can have strong effects on the inferred
mass distribution (\citealt{bin82}). To obtain the most accurate
results it is therefore important to make models that are as general
as possible. Of particular importance for collisionless, unrelaxed
systems such as galaxies is to constrain the velocity anisotropy using
available data, rather than to assume it a priori.

In a collisionless system the distribution function satisfies the
collisionless Boltzmann equation. Analytical methods to find solutions
of this equation usually rely on the Jeans Theorem, which states that
the distribution function must depend on the phase-space coordinates
through integrals of motion (quantities that are conserved along a
stellar orbit). In a spherical system all integrals are known
analytically, namely, the energy $E$ and the components of the angular
momentum vector ${\vec L}$. Analytical models for spherical systems
are therefore fairly easily constructed. In an axisymmetric system
things are more complicated (e.g., \citealt{bt87,mer99}). Only two
integrals are known analytically, $E$ and the vertical component
$L_{\rm z}$ of the angular momentum vector\footnote{We adopt the
notation in which $(x,y,z)$ denote the coordinates intrinsic to the
axisymmetric stellar system, with the plane $x-y$ being the equatorial
plane, and $z$ the symmetry axis. These relate via the inclination $i$
to the observable coordinates $(x',y')$ on the plane of the sky
(aligned, respectively, along the projected major and minor axes of
the stellar system), and $z'$ the line-of-sight direction, positive in
the direction away from us.}, but there is generally a third integral
for which no analytical expression exists. Therefore, it is not
generally possible to construct an axisymmetric model
analytically. The special class of so-called `two-integral'
($f=f(E,L_z)$) models (e.g., \citealt{bat93,deh94,ver02}) has its uses
(e.g., \citealt{mag98,vdm06}), but these have an isotropic velocity
distribution in their meridional plane, which need not be a good fit
to real dynamical systems.

The most practical way to model a general axisymmetric system is to do
it numerically. While a few methods exist to do this (e.g.,
\citealt{m2m,nmagic}), the most common approach uses Schwarzschild's
(1979) method. One starts with a trial guess for the gravitational
potential $\Psi$ and then numerically calculates an orbit library that
samples integral space in some complete and uniform way. The orbits
are integrated for several hundred orbital periods, and the
time-averaged intrinsic and projected properties (density, LOS
velocity, etc.) are stored as the integration progresses. The
construction of a model consists of finding a weighted superposition
of the orbits that: (1) reproduces the observed stellar or surface
brightness distribution on the sky; and (2) reproduces all available
kinematical data to within the observational error bars. Additional
constraints can be added to enforce that the distribution function in
phase space be smooth and reasonably well behaved, e.g., through
regularization or by requiring maximum entropy.

Several axisymmetric Schwarzschild codes have been developed in the
last decade (e.g., \citealt{vdm98,cre99,geb00,val04,tho04}). These
codes deal with the situation in which information on the
line-of-sight velocity distribution (LOSVD) is available for a set of
positions on the projected plane of the sky. This is the case, e.g.,
when the kinematical data are from long-slit or integral-field
spectroscopic observations of unresolved galaxies. The optimization
problem for such data can be reduced to a linear matrix equation for
which one needs to find the least-squares solution with non-negative
weights \citep{rix97}. One dimension of the matrix corresponds to the
number of orbits in the library, while the other corresponds to the
number of (luminosity, kinematical and regularization) constraints
that must be reproduced. Both dimensions are typically in the range
$10^3$--$10^4$. Nonetheless, efficient numerical algorithms exist to
find the solution, which yield the orbital and the velocity
distribution of the model, as well as the $\chi^2$ of the fit to the
kinematical data. The procedure must then be iterated with different
gravitational potentials, to determine the potential that provides the
overall best $\chi^2$. The existing codes have been used and tested
extensively (e.g.,
\citealt{cre00,cap02,cap06,geb03,ben05,dav06}). Some questions remain,
e.g., about the importance of smoothing in phase space, the exact
meaning of the confidence regions determined using $\Delta \chi^2$
contours, and, in some situations, valid concerns have been raised
regarding whether the available data contain enough information so as
to warrant the conclusions of the \sch modeling
\citep{val04,cre04,kra05}. Nevertheless, on the whole Schwarzschild
codes have now been established as an accurate and versatile tool to
study a wide range of dynamical problems.

A disadvantage of the existing codes is that they cannot be easily
applied to the large class of problems in which the kinematical
observations come in the form of discrete velocity measurements,
rather than as LOSVDs. This is encountered, e.g., when modeling the
dynamics of galaxies at large radii, where the low-surface brightness
prevents integrated-light spectroscopy. The only available data are
then often of a discrete nature, e.g., via the LOS velocities of
individual stars in galaxies of the Local Group (e.g.,
\citealt{bill00,jan01,jan02,lok02,wil04,lok05,wal06,geh06}), or via
planetary nebulae (e.g., \citealt{dou02,rom03,teo05}) and globular
clusters (e.g., \citealt{cote01,tom04}) surrounding giant
ellipticals. The kinematical data available for clusters of galaxies,
consisting of redshifts for individual galaxies, are of a similarly
discrete nature (e.g., \citealt{lok03}). The typical datasets in all
these cases consist of tens to hundreds of LOS velocities. Galactic
globular clusters constitute another class of object for which
kinematical data is often available only as discrete measurements,
rather than in the form of LOSVDs. From ground-based observations,
data sets of individual LOS velocities can be available for up to
thousands of stars in these systems (e.g.,
\citealt{sun96,may97,rei06}), and for $\omega$ Cen it has been
possible to assemble large samples of proper motions as well
\citep{vleu00}. With the capabilities of {\it HST}, accurate proper
motion data sets with up to $\sim 10^4$ stars are now becoming
available for several more Galactic globular clusters (e.g.,
\citealt{mcn03,mcl06}). 

Note that discrete datasets do not necessarily provide better or worse
information than datasets obtained from integrated-light
measurements. Both types of data have their advantages and
disadvantages. For discrete datasets, for example, interloper
contamination can be a problem (see also the end of
Section~\ref{sec:logL} below). By contrast, for integrated-light
measurements, it is often difficult to constrain the wings of the
LOSVD due to uncertainties associated with continuum
subtraction. Which type of data is most appropriate and most easily
obtained depends on the specific object under study. This is therefore
not a question that we address in this paper.  Instead, we focus on
the issue of how to best analyze discrete data, if that happens to be
what is available.

Analyses of discrete datasets have often been more simplified than the
analyses that are now common for integrated-light data. For example,
the observations are analyzed using the Jeans equations (e.g.,
\citealt{ger02,lok03,cote03,dou07}), often with the help of data
binning to calculate rotation velocity and velocity dispersion
profiles (see, however, the ``spherical'' Schwarzschild models of M87
of \citealt{rom01}). The disadvantage of such an approach is that not
all the information content of the data is used, including information
on deviations of the velocity histograms from a Gaussian. Such
deviations are important because they constrain the velocity
dispersion anisotropy of the system (e.g.,
\citealt{vdm93,ger93,ger98}). This anisotropy is an important
ingredient in some existing controversies, e.g. regarding the presence
of dark halos around elliptical galaxies \citep{rom03,dek05}. Loss of
information can be avoided when large numbers of datapoints are
available, as is often the case for globular clusters. It is then
possible to create velocity histograms for binned areas on the
projected plane of the sky, after which analysis can be done with
existing Schwarzschild codes (e.g., \citealt{bos06}). While this is
possible for large datasets, such an approach is not viable for the
more typical, smaller datasets that are often available. The
availability of Schwarzschild codes that can fully exploit the
information content of such smaller datasets would therefore be
valuable to advance this subject.

Motivated by these considerations we set out to adapt our existing
Schwarzschild code \citep{vdm98} to deal with discrete datasets. This
does not constitute a trivial change, since it changes the constrained
superposition procedure from a linear matrix problem to a more
complicated maximum likelihood one. For each observed velocity of a
particle in the system the question becomes: what is the probability
that this velocity would have been observed if the model is correct?
The overall likelihood of the data, given a trial model, is the
product of these probabilities for all observations. Such likelihood
problems have previously been solved for spherical systems
\citep{mer93,vdm00,wu06} and the special class of axisymmetric
$f(E,L_z)$ systems \citep{mer97,wu07}. However, for the axisymmetric
Schwarzschild modeling approach the problem corresponds to finding the
minimum of a function in a space with a dimension of
$10^3$--$10^4$. We show in this work, via the \sch modeling of
simulated datasets, that this problem can indeed be solved
successfully and efficiently. Moreover, we follow \cite{glenn06} and
implement in our new code the ability to calculate and fit proper
motions in addition to LOS velocities. Applications of the code to
real datasets will be presented in forthcoming papers.

The structure of the paper is as follows. In Section \ref{sec:logL} we
phrase the new problem of fitting a \sch model to a dataset of
discrete velocities (of one, two, or three dimensions) of individual
kinematic tracers in terms of a likelihood formalism. Section
\ref{sec:code} describes the implementation of the discrete fitting
procedure into our existing \sch code. At the same time, we summarize
here the major steps involved in the construction of the probability
matrix that describes the likelihood of a given kinematic data point
belonging to some particular orbit of the library. We then present in
Section \ref{sec:tools} sets of simulated data that we use for the
purpose of testing the performance of the discrete \sch code. We also
describe the known input distribution functions from which these data
were drawn. The application of the code to the simulated datasets is
presented in Section \ref{sec:tests}. We present a thorough analysis
of the accuracy with which our discrete \sch code recovers the known
distribution function, mass-to-light ratio and inclination used to
generate the simulated data. Finally, in Section \ref{sec:end} we
summarize our findings and present our conclusions.

\section{Linear and non-linear constraints in the likelihood formalism}
\label{sec:logL}

In the \sch scheme the properties of every orbit $j$ in the orbit
library are computed and stored. The modeling consists in finding the
superposition of orbital weights $a_j^2$, i.e., the fraction of
particles in the system residing in each orbit, that best reproduces
some set of constraints. The weights are written as squares to ensure
that they never become negative. Linear constraints are of the form
\begin{equation}
\label{eq.constraint}
  {\gamma_k}^* = {\gamma_k} \pm {\sigma_k} , \quad k=1 \ldots M .
\end{equation}
Here ${\gamma_k}$ is a constraint value that needs to be reproduced,
${\sigma_k}$ is its uncertainty, and ${\gamma_k}^*$ is its model
prediction
\begin{equation}
\label{eq.gamma.def}
  {\gamma_k}^* = \sum_j B_{kj} a_j^2/\sum_j a_j^2 .
\end{equation}
The matrix $B_{kj}$ represents here, for orbit $j$, the probability
distribution corresponding to the constraint $\gamma_k$. The
constraints are generally one of the following: (a) the integrated
light (surface brightness) of a stellar population in some aperture
number in the projected plane of the sky, necessary to reproduce an
observational measurement of the surface brightness; (b) the mean LOS
velocity, velocity dispersion, or for data of sufficient quality, a
higher-order Gauss-Hermite moment in some aperture number in the
projected plane of the sky, necessary to reproduce an observational
measurement of the stellar kinematics; (c) the integrated mass in some
meridional $(R,z)$ plane grid point, necessary to provide a consistent
model; (d) a combination of distribution function moments in some
meridional $(R,z)$ plane grid point, if a model with a particular
dynamical structure is desired (e.g., one may want a model with $\rho
(\overline{v_R^2}-\overline{v_z^2})$ equal to zero in order to
simulate a two-integral $f(E,L_z)$ model); (e) a combination of orbit
weights, if regularization constraints are desired to enforce
smoothness of the model in phase space (e.g., one can set the N-th
order divided difference of adjacent orbit weights to zero, with an
uncertainty $\Delta {\gamma_k}$ that measures the desired amount of
smoothing).

It is natural to choose the best-fitting model to be the one that
produces the maximum likelihood. To determine the likelihood we need
to write down an expression for the probability of measuring
$\gamma_k$ among all its possible values. To do this, we recall that
any model is not an attempt to reproduce a set of observations to
infinite accuracy, but instead to do it within the uncertainty
$\sigma_k$. For observational constraints, such as those in (a) and
(b) above, $\sigma_k$ is equal to the measurement uncertainty. For
other constraints, such as those in (c)-(e) above, $\sigma_k$ can be
used as a forcing parameter that compels how accurately the likelihood
needs to peak around a particular value of $\gamma_k$. If one assumes
that these uncertainties have a normal (Gaussian) distribution, then
the probability we are interested in is given by
\begin{eqnarray}
\label{eq.non.linear.term}
P(\gamma_k) &=& 
{1\over \sqrt{2\pi} {\sigma_k}} 
\exp\left[-{1\over 2 {\sigma_k}^2}
\left(\gamma_k- {\gamma_k}^* \right)^2 \right] .
%\\
%&=&{1\over \sqrt{2\pi} {\sigma_k}} 
%\exp\left[-{1\over 2 {\sigma_k}^2}
%\left(\gamma_k- {\sum_j  B_{kj} a_j^2 \over\sum_j a_j^2} \right)^2 
%\right] . \nonumber
\end{eqnarray}
The combined probability for the simultaneous occurrence of all $M$
linear constraints is then given by the product of the single
probabilities, $L_{\rm linear} = \prod_k P(\gamma_k)$. Using equation
(\ref{eq.non.linear.term}), the logarithm of this linear part of the
likelihood is therefore

\begin{equation}
\label{eq:loglinear}
  \ln L_{\rm linear} = - \sum_{k=1}^M  \ln {\sqrt{2\pi} \sigma_k} \, - \,
          \sum_{k=1}^M \left(
                  \frac{\gamma_k-\gamma_k^*}{\sqrt{2}\,\sigma_k}\right)^2 . 
\end{equation}
The first sum on the right-hand side of this expression does not
depend on the orbital weights $a_j^2$ and, therefore, does not affect
the likelihood maximization. The second term has the exact form of the
$\chi^2$ statistic. Maximizing the likelihood is therefore equivalent
to the minimization of this $\chi^2$. This can be done by finding the
solution of the set of equations~(\ref{eq.constraint}) and
(\ref{eq.gamma.def}), which can be rewritten as an overdetermined
matrix equation. This matrix equation can be solved with the use of
standard non-negative least-squares (NNLS) algorithms (see
\citealt{rix97} for a detailed description).

In the case of discrete data, however, the introduction of constraints
of a ``non-linear'' type is inevitable in order to adequately exploit
the entire information content available, avoiding restrictive
simplifications and loss of information due to binning.  This occurs
because the individual probabilities do not necessarily have the
simple, Gaussian form of equation (\ref{eq.non.linear.term}).  The
procedure for finding the maximum likelihood then cannot be cast as
the solution of a linear matrix equation anymore.

Suppose we have a kinematic dataset consisting of discrete
measurements which we are trying to model using the \sch
technique. Let $P_j({\bf q})$ be the phase-space probability
distribution of any given orbit, properly averaged azimuthally, and
normalized such that $\int{P_j({\bf q}){\rm d^3}r\,{\rm d^3}v} = 1$.
We use ${\bf q}$ to denote a vector of up to six Euclidean spatial and
velocity coordinates. Whenever ${\bf q}$ is shorter than 6 elements,
it is understood that the distribution has been marginalized over the
missing dimensions. Then the total probability of drawing a particle
from a superposition of orbits representing the whole system is
\begin{equation}
\label{eq.prob.q}
  P({\bf q}) = \sum_j a^2_j P_j({\bf q}) /\sum_j a_j^2 .
\end{equation}

We now need to consider the total probability of the ensemble of $N$
particles with kinematic information that constitute our discrete
dataset. Before this, however, it is necessary to make the
distinction, in the language of probabilities, between the possible
modes of sampling of the tracers available in a system of particles.
The two main possibilities depend on whether the particles are
randomly or non-randomly drawn from their spatial distribution, and we
may refer to these, respectively, as random positional sampling and
incomplete positional sampling. Additionally, particles may be drawn
with or without velocity information, thus adding up to a total of
four possibilities. The case with incomplete positional sampling and
no velocity information, however, does not provide any useful
constraint to the analysis and therefore we restrict the discussion to
the remaining three cases.

For particles drawn randomly from the spatial distribution with no
velocity information, the probability $P({\bf q})$ is

\begin{equation}
\label{eq.prob.r}
P({\bf q}) = P({\bf r}) = \sum_j a^2_j P_j({\bf r}) /\sum_j a_j^2\,,
\end{equation}

\noindent where {\bf r} represents a 2 or 3 dimensional position. This
type of dataset could result from imaging of the resolved populations
of a stellar system, where the positional information could be used as
actual constraints. This would force the model to fit the underlying
spatial distribution of discrete tracers, instead of making use of a
parametrization of the (continuous) brightness profile of the system.
Of course, a dataset without velocity information cannot by itself
constrain the dynamical state or the mass of the system.

In the case of random positional sampling including velocity
information, particles are randomly drawn from both the spatial and
velocity distributions. In this case, $P({\bf q})$ has the form

\begin{equation}
\label{eq.prob.rv}
P({\bf q}) = P({\bf r},{\bf v}) = \sum_j a^2_j P_j({\bf r},{\bf v}) /\sum_j a_j^2\,,
\end{equation}

\noindent where ${\bf r}$ is the same as above and ${\bf v}$
represents a general 1, 2, or 3 dimensional velocity. This would be
the case when being able to obtain the velocities of particles in a
given field without introducing any spatial or velocity bias, such as
the proper motions of all stars (brighter than some magnitude limit)
in a sufficiently sparse stellar cluster, or when LOS velocities are
obtained for a complete (or possibly magnitude-limited) set of
globular clusters or planetary nebulae in a galaxy.

In contrast, having {\it incomplete} positional sampling means that
the particles are drawn from a velocity distribution, with {\it a
priori} fixed positions. This can occur, for example, when because of
the usually limited availability of telescope time and resources, LOS
velocities are measured only for stars within some distance from the
photometric major or minor axes of a galaxy, or when because of the
finite size of fibers in a fiber-fed spectrograph, not all the
potentially observable kinematic tracers in the field can be actually
acquired. Incomplete positional sampling also arises when, even though
particles can be randomly drawn spatially, this is the case only for a
limited area. This occurs, for example, when the observations have to
avoid the innermost regions of a galaxy or globular cluster, where,
because of crowding, stars cannot be individually resolved. In these
case, $P({\bf q})$ has the form

\begin{equation}
\label{eq.prob.rfix}
P({\bf q}) = P({\bf v}|{\bf r}) = \sum_j a^2_j P_j({\bf r}) P_j({\bf v}|{\bf r})
    /\sum_j a_j^2 P_j({\bf r}),
\end{equation}

\noindent where, rather than just $a_j^2$, the effective weights when
summing together the individual orbital distributions are
$a_j^2\,P_j({\bf r})$.

Once the individual probabilities for all possible cases of spatial
sampling that comprise the data have been properly assigned, we can
proceed to the construction of the total probability of observing the
entire dataset. Let $N_1$ and $N_2$ be the number of observational
data points obtained under the mode of random positional sampling
without and with velocity information, respectively, and $N_3$ the
number of data points obtained with incomplete positional sampling
with kinematic information. Then, the total probability is simply the
product of the individual probabilities, with logarithm given by $\ln
L_{\rm discrete} = \sum_{i=1}^{N_1}\ln P({\bf r_i}) +
\sum_{i=1}^{N_2}\ln P({\bf r_i},{\bf v_i}) + \sum_{i=1}^{N_3}\ln
P({\bf v_i}|{\bf r_i})$. Using equations (\ref{eq.prob.r}) to
(\ref{eq.prob.rfix}), and adopting the abbreviated notation
$p^{(r)}_{i,j} \equiv P_j({\bf r_i})$, $p^{(q)}_{i,j} \equiv P_j({\bf
r_i,v_i})$, and $p^{(v)}_{i,j} \equiv P_j({\bf v_i | r_i})$ (all known
for each orbit $j$ and particle $i$ from the orbit library
calculation; see \S\,3), the quantity to maximize becomes
\begin{eqnarray}
\label{eq.total.likelihood.1}
\ln L_{\rm discrete}
&=&  \sum_{i=1}^{N_1} \left(\ln \sum_j a^2_j p^{(r)}_{i,j} -  \ln \sum_j a^2_j\right) \\
&+&  \sum_{i=1}^{N_2} \left(\ln \sum_j a^2_j p^{(q)}_{i,j} -  \ln \sum_j a^2_j\right) \nonumber \\
&+&  \sum_{i=1}^{N_3} \left( \ln \sum_j a^2_j p^{(r)}_{i,j}  p^{(v)}_{i,j} -  \ln \sum_j a^2_j  p^{(r)}_{i,j} \right). \nonumber
\end{eqnarray}
Joining the results in equations (\ref{eq:loglinear}) and
(\ref{eq.total.likelihood.1}), the complete log-likelihood for a
general application of the \sch method, which is the full expression
to be maximized with respect to the orbital weights $a_j$, is the sum
of the log-likelihoods for linear and discrete constraints
\begin{eqnarray}
\label{eq.logL}
\ln L = \ln L_{\rm linear} + \ln L_{\rm discrete}.
\end{eqnarray}
Finding the maximum likelihood corresponds to finding the solution of
$\partial(\ln L)/\partial a_l$ = 0, for all $l$. Denoting $s=\sum_j
a_j^2$, the expression for the first derivative is
\begin{eqnarray}
\label{eq.1st.deriv}
{\partial \ln L \over \partial a_l} 
&=& -\,{2 a_l \over s}
\sum_{k=1}^M {1\over {\sigma_k}^2} 
       \left( \gamma_k - {\gamma_k}^* \right)
       \left( - B_{kl} + {{\gamma_k^*}}   \right)\\
&& +\,2 a_l \sum_{i=1}^{N_1} \left( {p^{(r)}_{i,l}\over \sum_j  a^2_j p^{(r)}_{i,j}} - {1\over s} \right)\nonumber \\
&& +\,2 a_l \sum_{i=1}^{N_2} \left( {p^{(q)}_{i,l}\over \sum_j  a^2_j p^{(q)}_{i,j}} - {1\over s} \right)\nonumber \\
&& +\,2 a_l \sum_{i=1}^{N_3} \left( {p^{(r)}_{i,l} p^{(v)}_{i,l}\over 
     \sum_j  a^2_j p^{(r)}_{i,j}  p^{(v)}_{i,j}}
  -  { p^{(r)}_{i,l} \over \sum_j  a^2_jp^{(r)}_{i,j} } \right) . \nonumber
\end{eqnarray} 

One important question that remains is regarding the estimation of
confidence regions around the parameters of the best-fitting model,
i.e., the (statistical) uncertainties around the likelihood maximum in
the case of non-linear constraints. Recalling that maximizing $\ln L$
is equivalent to minimizing the quantity $\lambda = -2\ln L$, it is
easy to realize that, if the probabilities involved in equation
(\ref{eq.total.likelihood.1}) were all of Gaussian form, then
$\lambda$ would simply reduce to the well known $\chi^2$ statistic, as
we have already seen for the case with linear constraints in equation
(\ref{eq:loglinear}). When dealing with non-linear constraints,
however, the likelihood does not reduce to a simple $\chi^2$
form. Nevertheless, one still can use another well known theorem of
statistics which, used before by \citet{mer93a} and \citet{vdm00},
states that the ``likelihood-ratio'' statistic $\lambda - \lambda_{\rm
min}$ does tend to a $\chi^2$ statistic in the limit of large $N$,
with the number of degrees of freedom equal to the number of free
parameters that have not yet been varied and chosen so as to optimize
the fit. Therefore, the likelihood-ratio statistic reduces to the
$\Delta\chi^2$ statistic for $N\rightarrow\infty$, even though the
probabilities in equation (\ref{eq.total.likelihood.1}) are not all
individually Gaussian. Since in the present work we explore datasets
consisting of 100 kinematic measurements or more, the condition of
large $N$ should be reasonably fulfilled. Therefore, following the
likelihood-ratio statistic, we assume $\Delta\chi^2 = -2(\ln L-\ln
L_{\rm max})$, and compute the confidence regions around the best-fit
parameters in the usual way (e.g., \citealt{recipes}), i.e., with the
$1\sigma$ error for a single parameter corresponding to wherever
$\Delta\chi^2 = 1$, and so forth. Other approaches to quantify the
uncertainties exist as well, e.g., using Bayesian statistics, but
these are generally more difficult to implement (e.g.,
\citealt{mag06}).

The equations described above assume that any possible ``interloper''
contaminants have already been removed, and that the targets with
observed velocities that enter the likelihood equations all belong to
the system under study. For realistic datasets, contamination by
interlopers can certainly be a problem \citep{lok05}; i.e., targets
that happen to lie close to the line-of-sight of the stellar system
under study and are difficult to reject from the sample. However,
efficient interloper rejection schemes do exist for various types of
samples and these have been well-described in the literature
\citep{woj07a,woj07b}. Moreover, the use of empirically-calibrated
selection criteria (independent of the measured velocity) can produce
extraordinarily clean samples for kinematic analysis
\citep{gil06,gil07,sim07}. Either way, interloper rejection is best
discussed in the context of specific data sets. We therefore do not
discuss it further in the present paper. Interloper rejection for
discrete data sets can also be built in as part of the likelihood
analysis \citep{vdm00}, so a simple modification of the likelihood
equations given above could deal with interlopers explicitly. However,
we have not yet explored this in the present context.

\section{Computational Implementation}
\label{sec:code}

Given equations (\ref{eq.logL}) and (\ref{eq.1st.deriv}), fitting a
\sch model to the data requires the following two steps: (a)
determination of all the individual probabilities $p_{i,j}$ and matrix
elements $B_{kj}$, so that the only unknowns in equation
(\ref{eq.1st.deriv}) are the coefficients $a_l$; and (b) performing
the maximization of the total likelihood, i.e., finding the set of
orbital weights $a_l$ that satisfies $\partial(\ln L)/\partial a_l$ =
0, for all $l$, and therefore best fits the available constraints. The
elements of the matrix $B_{kj}$, corresponding to the linear
constraints discussed in \S\,\ref{sec:logL}, are calculated in the
same way as in the old (continuous) implementation of the code, and
for them we refer to \citet{rix97}, \citet{vdm98} and
\citet{cre99}. In what follows we concentrate on the probabilities
$p_{i,j}$ associated with the discrete treatment that is the subject
of this work.

\subsection{Calculation of Individual Probabilities}
\label{sec:pij}

The matrix elements $p_{i,j}$ in equation (\ref{eq.1st.deriv}), which
keep track of the probability that orbit $j$ of the library would have
produced the measurement $i$ of the dataset (each $j$ corresponding to
some combination of the three integrals of motion $E$, $L_z$, and
$I_3$), are stored as the orbit in question is being computed. That
is, at every time step during the orbit integration, we check whether
the position and velocity along the orbit is consistent with any of
the observational datapoints.  To accomplish this, it is necessary to
implement some degree of {\it smoothing}, both in position and
velocity space, since otherwise the probability of having a particle
on an orbit at exactly the observed position and velocity would be
infinitesimally small.

Smoothing in the spatial coordinates is accomplished through the
definition of an {\it aperture} around the position of each particle
in the dataset, with the size of the aperture controlling the amount
of smoothing. The optimal aperture size will be somewhat dependent on
the sampling characteristics of the data. In general, apertures should
not be too small, or otherwise few time steps during orbit integration
will fall on any one of them. This would lead to large shot noise in
the computed probabilities $p_{i,j}$, unless the orbits are integrated
for very long times. Nor should the apertures be too large, so that
information on the orbital structure of the model is not erased by
excessive spatial smoothing. The choice of aperture shape is arbitrary
and a matter of numerical convenience. We adopt square apertures as in
previous implementations of the code (long-slit observations naturally
produce data for rectangular apertures), and set their sizes to a
user-supplied fraction of $R$, the radius in the projected plane at
the aperture's position.

Once the spatial apertures are defined, and every time the projected
position along the orbit being integrated falls within an aperture, we
need to keep track of whether the orbital velocity matches the
observed velocity. In the old (continuous) implementation, the LOSVD
was computed and stored for every orbit $j$ at each aperture $i$, with
the size of the bins in the histogram determining the amount of
smoothing in velocity space. In our discrete treatment of the problem,
$p_{i,j}$ would simply be the histogram value for the bin that
contains the observed velocity. A direct, though information ally
incomplete, generalization of this implementation to kinematical data
in three-dimensions would be to keep track of two additional
histograms at each aperture to account for $\mu_{x'}$ and
$\mu_{y'}$. This has been done by \citet{glenn06} and \citet{bos06},
who calculated moments of the three model velocity distributions and
fitted them to those obtained from binning LOS and proper-motion
observations of stars in $\omega$ Cen and M15, respectively (note that
these studies still handle the data in a continuous fashion, by
reducing the initially discrete datasets to binned velocity
distributions at a number of apertures on the sky, an approach only
possible thanks to the very large number of stars with measured
velocities in these systems).

While reproducing the three-dimensional mean velocities and
dispersions of the stars in a stellar system is already an improvement
over all previous implementations of the \sch technique, doing so is
nevertheless a simplification of the problem. The reason is that it
implicitly assumes that the three velocity components are independent
of each other, i.e., it does not account for the fact that there is a
velocity ellipsoid whose cross terms are, in the most general case,
not identical to zero. The most complete treatment would be to store a
cube with entries for all possible combinations of
$(\mu_{x'},\mu_{y'},v_{z'})$, and do this at each spatial aperture where there
is kinematical data available.  This implementation would be, however,
expensive in terms of memory storage and, moreover, not absolutely
necessary, simply because we are not interested in the entire
probability cube. Instead, we only need probabilities in the cases
when the model velocities are close to the observed ones. Thus, in the
framework of velocity histograms or full velocity cubes, and because
of the discrete nature of the data, the large majority of the bins or
entries would be filled with weights that do not affect the likelihood
in equation (\ref{eq.logL}).

Therefore, we adopt an approach in which, instead of storing velocity
histograms or cubes, every time an orbit $j$ passes through an
aperture $i$ with kinematical data, we add a Gaussian contribution to
$p_{i,j}$. This contribution is centered on the observed
(any-dimensional) velocity and has a dispersion that reflects the
measurement errors, and if desired, any amount of extra velocity
smoothing. Thus, denoting the actually measured components of the
particle's velocity in aperture $i$ as $v_{ik}$ and their associated
uncertainties as $e_{ik}$, with $k=1\ldots3$ corresponding to
$v_{x'}$, $v_{y'}$, and $v_{z'} = v_{\rm los}$, the multiplicative
contribution $w_{ik}^{(j)}$ to the probability has the form
\begin{eqnarray}
\label{eq.weights}
w_{ik}^{(j)} &=& 
{1\over \sqrt{2\pi\left(\xi_k^2 + e_{ik}^2\right)}} \exp{\left[-\,\frac{\left(v_{jk}-v_{ik}\right)^2}{2 \left(\xi_k^2+e_{ik}^2\right)}\right]},
\end{eqnarray}
where $v_{jk}$ is the component $k$ of the velocity of a test particle
on orbit $j$. The quantity $\xi_k$ is the numerical smoothing assigned
to velocity component $k$. Whenever a particular component $k$ of the
velocity is not available, we set $w_{ik}^{(j)} = 1$. Finally, in
order to account for the fact that we represent a continuous orbit by
a discrete sequence of time steps, we weigh this Gaussian factor by
multiplying it by the timestep $\Delta t_j$. Therefore, for every
orbit $j$, and every time the orbit integration falls within an
aperture, the probability is increased according to
\begin{eqnarray}
\label{eq.pij}
p_{i,j} = p_{i,j} + \Delta t_j \prod_{k=1}^3 w_{ik}^{(j)}.
\end{eqnarray}
When the integration of orbit $j$ is done, the $p_{i,j}$ elements for
all datapoints (apertures) are written to a file for later use by the
algorithm that performs the maximization of the likelihood. 

In most practical applications one can set $\xi_k = 0$, since the
error bars $e_{ik}$ on the data already provide sufficient natural
smoothing for numerical efficiency. We do this throughout the rest of
this paper. However, we note that there may be situations in which
non-zero $\xi_k$ may be beneficial. For example, if the observational
errors $e_{ik}$ are much smaller than the velocity dispersions
$\sigma_k$ of the system. It then takes very long integrations to beat
down the shot noise in the orbital distributions $p_{i,j}$. Addition
of a numerical smoothing $\xi_k$ with $e_{ik} \ll \xi_k \ll \sigma_k$
can then speed up the calculations without affecting the accuracy of
the results.
 
The approach of equations (\ref{eq.weights}) and (\ref{eq.pij})
assumes that the errors $e_{ik}$ for the different datapoints are
uncorrelated. Sometimes this is not true, as in the case of the proper
motions of stars in the globular cluster $\omega$ Cen, where relative
rotation between the old photographic plates used in their derivation
produce an artifact overall rotation of the cluster
\citep{glenn06}. If problems like these can not be removed before
modeling, a more complicated treatment than the one described here
will be necessary.

\subsection{Finding the maximum likelihood solution}
\label{sec.mkfitin}

The non-linear nature of the discrete problem addressed in this paper
requires the use of a non-linear optimizer, and there is no guarantee
of a unique optimum. After experimentation with various optimization
algorithms, we settled on the TOMS 500 conjugate gradient optimizer of
\citet{sha80}. This code uses the function value and gradient to
optimize along successive vectors (lines) in the space of the orbital
weights, choosing the optimization direction at every pass in a manner
that attempts to minimize the number of such line minimizations needed
(see Chapter 10 of \citealt{recipes} for details on conjugate gradient
methods).

In our code, we rely on the fact that the majority of orbits do not
contribute to any particular linear constraint, or to the likelihood
of any particular observational datum. In the notation of equation
(\ref{eq.1st.deriv}), the linear constraints $B_{k,l}$, and also the
$p_{i,l}$, are sparse matrices. Accordingly, the code to evaluate the
gradient in equation (\ref{eq.1st.deriv}) is written to store and
evaluate only non-zero terms of $B_{k,l}$ and $p_{k,l}$, reducing the
computational burden by a factor of four or five.

\begin{figure}
%\epsscale{.80}
\plotone{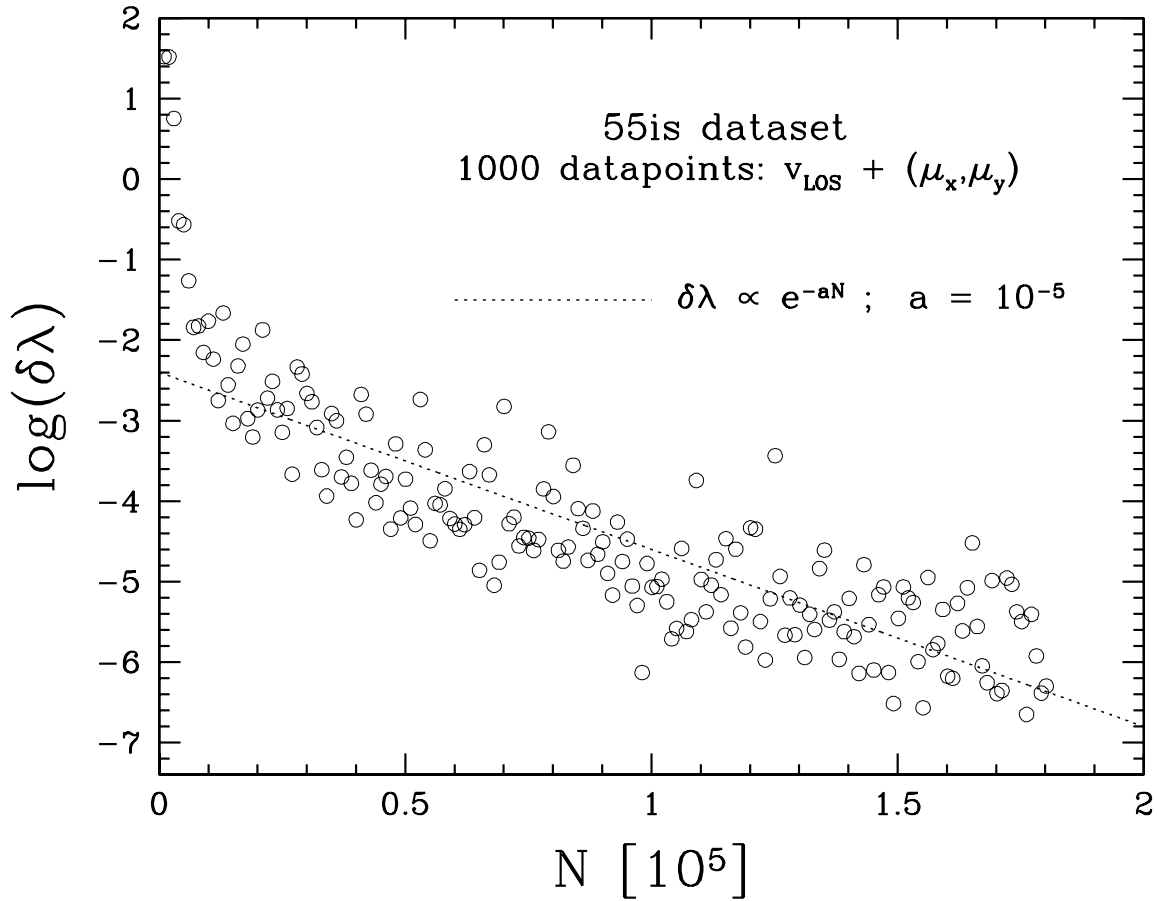}
\caption{\label{fig:mkfitin} Maximization of the total likelihood $\ln
L$ as a function of the number of function evaluations $N$, for a
typical \sch fit using a dataset consisting of 1000 discrete kinematic
measurements consisting of full 3-dimensional velocities (55is case;
\S\,\ref{sec:data} and Figure \ref{fig:data55is}). Shown on the
vertical axis is the change in the quantity $\lambda = -2\ln L$,
denoted as $\delta\lambda$. This change becomes smaller as the
optimization converges to a solution following approximately the
exponential relation illustrated by the dotted line. See discussion in
\S\,\ref{sec.mkfitin}. }
\end{figure}

To evaluate convergence and estimate the proximity of our final
likelihood maximum to the true (possibly local) maximum, we plot the
magnitude of the improvement of the likelihood $\delta\lambda$ as a
function of the number of function evaluations $N$. See Figure
\ref{fig:mkfitin}. We find that $\delta\lambda$ is well represented by
an exponential relation $\delta\lambda \sim \exp{(-aN)}$, where
$a\approx 10^{-5}$. Therefore, the future change in the likelihood if
the optimizer were allowed to run forever would be $\Delta \ln L \sim
\int_{N_0}^\infty \delta\lambda\, dN = a^{-1} (\delta\lambda)_0$,
where $(\delta\lambda)_0$ is the current change in likelihood at step
$N_0$. In practice, we terminate the optimization at $\delta\lambda =
10^{-6}-10^{-7}$, so that we expect to be within an additive factor of
$\leq 0.1$ of the true likelihood maximum. This typically occurs after
a number $N\sim 10^5$ of function evaluations.  The final accuracy is
merely linear in the exponential coefficient $a$, so that this
accuracy estimate should be reasonably robust.

We ran a variety of tests in order to establish whether the algorithm
has a tendency of finding local extrema as opposed to global ones. In
particular, for some of the test cases to be discussed later in
\S\,\ref{sec:tests}, we started the iterative algorithm from different
initial conditions, to verify that the solutions thus obtained were
always in (statistical) agreement. Also, as will be shown in
\S\,\ref{sec:tests}, we find that the algorithm recovers the
properties of known input models with reasonable accuracy. While this
does not prove that the \sch code cannot end up in a local maximum, at
least it shows that the code does not end up in (potential) local
maxima that are far from the correct solution.

In practice we usually start the maximization procedure from a
homogeneous set of initial mass weights. We also investigated whether
the convergence to a solution can be sped up by starting the iterative
process from initial conditions that may already be reasonably close
to the final solution. For example, we ran tests starting from a set
of weights corresponding to a two-integral DF of the form $f(E,L_z)$
that already fit the light (surface brightness) profile followed by
the input data. Such a solution is easily obtained as the NNLS
solution of a matrix equation. We found that the same final answer was
reached in essentially the same number of iterations.

%\newpage

\section{Pseudo-Data and Comparison Distribution Functions}
\label{sec:tools}

In order to test the performance of our discrete \sch code, we
generate sets of simulated data drawn from a known phase-space
distribution function (DF). Unlike the case of using actual
observations of a real stellar system, this approach offers the
advantage of unambiguously knowing in advance the input properties
underlying the data, which an optimally-working code should be able to
``recover''.  It also provides flexibility by allowing the possibility
of adapting the input data at will in order to test different aspects
of the code (\S\,\ref{sec:tests}). We discuss here the construction of
various sets of pseudo-data and the properties of the underlying
models.

\subsection{Simulated Datasets}
\label{sec:data}

Our simulated input data are obtained from a set of $f(E,L_z)$ DFs
derived by \citet{vdm98}, with the methodology for drawing N-body
initial conditions from these DFs described in \citet{vdm97b}. The
models have a constant mass-to-light ratio $\Upsilon$, and have
neither a central black hole or extended dark halo. They provide good
fits to available photometric and kinematic observations of the galaxy
M32 over the radial range from $1-20$ arcsec. However, this property
has no bearing on the present analysis. We only use the fact that
there is a known DF, and not that this DF resembles any realistic
stellar system. A two-integral $f(E,L_z)$ DF provides a useful test
case (see also \citealt{cre99}, \citealt{ver02}), and does not mean
that the model results would be less valid for more general DFs. Also,
the use of a constant $\Upsilon$ is motivated only to simplify the
test environment. Central black holes (e.g., \citealt{vdm98,geb00})
and extended dark halos (e.g., \citealt{rix97,cap06}) can be easily
implemented in any \sch code.

The luminous mass density is assumed to be axisymmetric and is
parameterized according to

\begin{equation}
\label{eq:light}
\rho(R,z) = \rho_{0}(m/b)^{\alpha}[1+(m/b)^2]^{\beta}[1+(m/c)^2]^{\gamma},
\end{equation}

\noindent with $m^2 \equiv R^2 + (z/q)^2$. Here, $q$ is the (constant)
intrinsic axis ratio, related to the projected (observed) axis ratio
$q_p$ via the inclination angle $i$, $q_p^2 = \cos^2 i + q^2\sin^2
i$. The parameters in equation (\ref{eq:light}) are set to
$\alpha=-1.435$, $\beta=-0.423$, $\gamma=-1.298$, $b=0.\arcsec55$,
$c=102.\arcsec0$, $q_p=0.73$, and $\rho_0=j_{0}\Upsilon_0$, with the
$V$-band luminosity density $j_0 =
0.463\times10^5(q_p/q)$\lsun\,pc$^{-3}$, and $\Upsilon_0$ the
mass-to-light ratio in the $V$-band and in solar units. The adopted
distance is 0.7 Mpc. The models share the property of appearing the
same in projection on the sky, but correspond to different intrinsic
axis ratios as determined by the inclination angle $i$.

The even part $f_e$ of the DF $f(E,L_z)$ is uniquely determined by the
mass density $\rho(R,z)$ (e.g., \citealt{bt87}). To specify the odd
part $f_o$ of the DF we follow \citet{vdm94} and write
\begin{equation}
\label{eq:odd}
f_o = f_e\,(2\eta-1)\,h_u[L_z/L_{z,{\rm max}}(E)],
\end{equation}
with $L_{z,{\rm max}}(E)$ being the angular momentum of a circular
orbit of energy $E$ in the equatorial plane ($z=0$), and the auxiliary
function $h_u$ defined by
\begin{equation}
\label{eq:ha}
h_u(x) = \left\{
 \begin{array}{ll}
    \tanh(ux/2)\,\,/\,\,\tanh(u/2)&  \mbox{$(u > 0)$},\\
    x&  \mbox{$(u=0)$},\\
    (2/u)\tanh^{-1}[x\tanh(u/2)]&  \mbox{$(u < 0)$}.
 \end{array}\right.
\end{equation}
The choice of the parameters $\eta$ and $u$ determines the degree of
streaming of the dataset. These free parameters can have values in the
ranges $0\leq \eta \leq 1$ and $-\infty < u < \infty$, with $\eta$
controlling the fraction of stars in the equatorial plane with
clock-wise rotation, and $u$ controlling the behavior of the stellar
streaming with orbital shape. The family of functions $h_u$ is shown
in Figure 1 of \citet{vdm94}. Combinations of $(\eta,u)$ that fit data
for M32 are also discussed in that paper. Here we explore a variety of
input datasets with different amounts of mean streaming and test the
recovery of these properties by our discrete \sch code.

We generated 6 different datasets to test our discrete \sch code. By
dataset we mean a number of particle $(x',y')$ positions on the sky
with corresponding proper motions $(\mu_{\rm x'},\mu_{\rm y'})$ and
LOS velocities $v_{\rm z'}$. For two chosen inclinations on the sky,
$i=90\grad$ and $i=55\grad$, we produced three datasets resembling
systems with varying degrees of rotation: a non-streaming system
($\eta=0.5$ and $u=1$), a maximally-streaming system ($\eta=1$ and
$u=\infty$), and a third system with intermediate streaming ($\eta=1$
and $u=0$). We label our different datasets as 90ns, 90is, and 90ms to
indicate the non-streaming, intermediate-streaming, and
maximally-streaming cases of $i=90\grad$, respectively. Similarly, for
the $i=55\grad$ case, we have the 55ns, 55is, and 55ms datasets. The
mass-to-light ratio used to generate the datasets is $\Upsilon_0=2.51$
for $i=90\grad$ and $\Upsilon_0=2.55$ for $i=55\grad$, in units of
\msun/$L_{\odot,V}$.

\begin{figure}
%\epsscale{.80}
%\plotone{3_maps.from_hoegaarden.10e4.ps}
\plotone{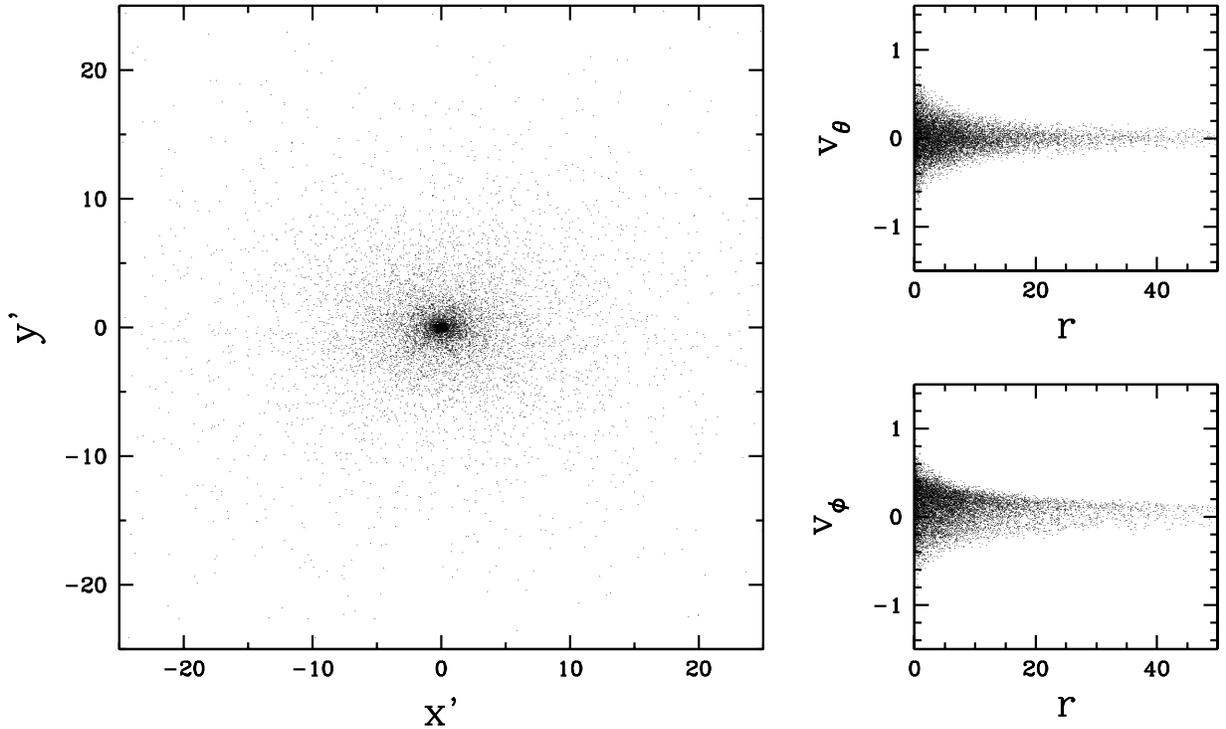}
\caption{\label{fig:data55is} Three phase-space projections using
10,000 particles of the 55is simulated dataset ($i=55\grad$ with
intermediate streaming). All input datasets have been constructed by
randomly drawing discrete particles from a two-integral distribution
function of the form $f(E,L_z)$, and are built so that, regardless of
their true inclination, they have the same light distribution when
projected on the plane of the sky. The coordinate system $(x',y',z')$
represents the observer's system, with $(x',y')$ on the plane of the
sky, and $z'$ the direction along the line-of-sight, defined positive
away from the observer. The coordinates $r$, $v_{\theta}$, and
$v_{\phi}$ correspond to the usual spherical coordinates intrinsic to
the system. Spatial coordinates are in units of 8.7 arcsec, and
velocities in units of 250 \kms. Note the asymmetry with respect to
$v_{\phi}=0$ in the bottom-right panel, reflecting the net rotation of
the 55is dataset. }
\end{figure}

Although we examined the performance of our \sch code with tests that
involve all of the six simulated datasets introduced above, we chose
to use the 55is dataset to show most of our results. Figure
\ref{fig:data55is} shows some projections of the phase-space
coordinates for the 55is dataset.

\subsection{Comparison DF}
\label{sec:DF}

In order to quantitatively judge the performance of the three-integral
\sch code, it is desirable to make a comparison between the properties
of the input DF (i.e., that from which the pseudo-data were obtained)
and those of the fitted DF (i.e., that found as the solution to the
fitting or minimization problem). It is important to note in this
context that the direct output of our \sch code is not in the form of
a proper DF $f$, but rather in the form of ``mass weights'' $\zeta$
associated to each set of integrals of motion $(E,L_z,I_3)$ that
uniquely define an orbit. The relation between the DF and the orbital
mass weights is through a volume element dependent on the three
integrals and an integration over the 3-dimensional space associated
to the particular orbit (see \citealt{voort84} for a detailed
discussion). Such a conversion can be done in Schwarzschild codes
(e.g., \citealt{tho04}), but this is not necessary for the goals of
the present paper. We therefore limit ourselves to the comparison
between the input and the fitted orbital mass weight distributions,
which from now on we denote by $\zeta_{\rm in}(E,L_z,I_3)$ and
$\zeta_{\rm fit}(E,L_z,I_3)$, respectively.

To validate the weights $\zeta_{\rm fit}(E,L_z,I_3)$ returned by the
\sch code, we need to know the weights $\zeta_{\rm in}(E,L_z,I_3)$ for
the model DF $f(E,L_z)$. This is not a simple problem in the absence
of an analytic expression for $I_3$. However, two related functions
are more easily accessible. The first is $\bar\zeta_{\rm in}(E,L_z)$,
defined as the projection of $\zeta_{\rm in}(E,L_z,I_3)$ over the
$E-L_z$ plane (i.e., integrated over $I_3$). Having the means of
drawing N-body initial conditions from the DF \citep{vdm97b}, we know
that the energy and z-component of the angular momentum of each
particle are given by $E=\psi - {1\over 2}v^2$ and $L_z = R\cdot
v_{\phi}$, respectively. Therefore, $\bar\zeta_{\rm in}(E,L_z)$ is
easily obtained by binning a large N-body dataset $(N\sim 10^6)$ in
$E$ and $L_z$. The second related function that is easily accessible
is $\zeta_{\rm Kep,\lambda}(E,L_z,I_3)$, the distribution of mass
weights for an $f(E,L_z)$ model of axial ratio $q$ and a power-law
density profile with logarithmic slope $\lambda$ in a spherical Kepler
potential. This function is calculated analytically in de Bruijne et
al. (1996; their equation (38)), and has the form
\begin{equation}
\label{eq:kep}
\zeta_{\rm Kep,\lambda}(E,L_z,I_3) = E^{\lambda-4}\times j_{\lambda}\left[L_z/L_{z,{\rm max}}(E),I_3\right].
\end{equation}
Here, $\lambda$ is the logarithmic slope of the mass distribution and
$j_{\lambda}$ is a known function. The spherical Kepler potential is
of course only an accurate approximation to our model at
asymptotically large radii. Nonetheless, we can combine
$\bar\zeta_{\rm in}(E,L_z)$ and $\zeta_{\rm Kep,\lambda}(E,L_z,I_3)$
to obtain a reasonable approximation for $\zeta_{\rm in}(E,L_z,I_3)$
throughout the system, namely
\begin{equation}
\label{eq:zetain}
\zeta_{\rm in}(E,L_z,I_3) \approx \bar\zeta_{\rm in}(E,L_z) \times g(I_3),
%j_{\lambda}\left(g(L_z,E),I_3\right).
\end{equation}
with
\begin{equation}
\label{eq:I3}
g(I_3) \equiv \frac{j_{\lambda}\left[L_z/L_{z,{\rm max}}(E),I_3\right]}{\int j_{\lambda}\left[L_z/L_{z,{\rm max}}(E),I_3\right]{\rm d} I_3}.
\end{equation}
For $\lambda$ we take the slope of the mass distribution of equation
(\ref{eq:light}) at $r=R_c$, the radius of the circular orbit of
energy $E$ in the equatorial plane $(z=0)$. The function $\zeta_{\rm
in}$ in equation (\ref{eq:zetain}) is correct (i.e., reduces to
$\bar\zeta_{\rm in}$) when projected on the $E-L_z$ plane, and has
approximately the correct distribution over $I_3$ at fixed
$(E,L_z)$. In this way, we construct sets of orbital mass weights for
each of our 6 simulated datasets described in \S\,\ref{sec:data}.

\section{Performance Tests}
\label{sec:tests}

Using all the kinematic (pseudo) datasets and their corresponding
input DFs described in \S\,\ref{sec:tools} we now proceed to examine
how accurately the discrete \sch code can recover the properties of
the galaxy models used to generate the input datasets. By {\it
recovery} we mean to determine how close or how far is the obtained
solution from the known DF, known mass-to-light ratio \ml, and known
inclination $i$ of the galaxy model corresponding to the simulated
dataset that was provided as input to the code. At the same time, we
investigate the reliability of the uncertainties provided by the code
on each of these properties.

In the general case of modeling real observations of an actual stellar
system, the true radial mass density profile is not known a priori and
is typically described following some parameterization. Since mass may
not necessarily follow light, or may do so in some complicated way,
different plausible mass models should be attempted, as well as
allowing for possible variations of the mass-to-light ratio with
position. For the purposes of the present tests, however, the
underlying mass distribution is assumed to be perfectly known from
equation (\ref{eq:light}), except for the value of \ml. Therefore, the
assumed parameterization for the mass distribution is only a
1-parameter family, and includes the ``correct'' distribution
$(\Upsilon = \Upsilon_0)$. In applications to real data,
higher-parameter families may be necessary, and there is no guarantee
that any member of the family would provide a good approximation to
the true underlying distribution.

The results of our tests are examined via three different exercises,
which can be performed on each of the 6 different input datasets,
providing a good baseline to judge the performance of our discrete
\sch code. First, we explore the recovery of the internal orbital
structure of the input dataset (i.e., the input DF, or more
specifically, the input mass weights $\zeta_{\rm in}$) by feeding the
code with the correct inclination and mass-to-light ratio \ml\,
(\S\,\ref{sec:getDF}). Second, we fix the inclination to the correct
value of the input dataset and study whether the code finds the
minimum of the $\Delta \chi^2$ function at the correct value of $\Upsilon$
(\S\,\ref{sec:getML}). And third, we explore grids of \sch models with
different ($i$,\ml) combinations to study how well these two
properties are recovered when they are both assumed unknown
(\S\,\ref{sec:grids}).

We run all the above exercises for various subsets of each of our 6
datasets in order to explore the results as a function of relevant
observational variables, particularly the size of the input dataset
and the type of kinematical constraints available (i.e., only LOS
velocities, only proper motions, or the complete three-dimensional
velocities). This adds even more elements for a thorough assessment of
the code's performance. It also provides insights into the types of
datasets that will be necessary to constrain $i$ or $\Upsilon$ to some
given uncertainty in realistic situations.

Our \sch code has the capability of computing and storing, during a
single orbit integration, the orbital properties for a series of
different values of \ml. Thus, during the construction of the orbit
library, different values of \ml\, are converted into a dimensionless
factor $v_{\rm s}$ that multiplies all our original velocities, thus
with \ml\, scaling simply as $v_{\rm s}^2$. We stress that this allows
us to explore several values of \ml\, while computing only one orbit
library. In our tests, we explore models for velocity factors in the
range $0.8\leq v_{\rm s} \leq 1.2$. Given that our galaxy models with
different inclinations have slightly different mass-to-light ratios
$\Upsilon_0$, the use of this dimensionless representation also
facilitates the visualization of the results in \S\,\ref{sec:getML}
and \S\,\ref{sec:grids}. The correct (input) value of \ml\, is always
at $v_{\rm s} = (\Upsilon/\Upsilon_0)^{1/2} = 1$.

%We start by discussing the ``standard'' parameter settings with which
%we have run most of our tests (\S\,\ref{sec:settings}). 

\subsection{Standard Settings}
\label{sec:settings}

At each of its different steps, the \sch code requires the user to
specify several settings (or dials) that control a corresponding
number of tasks and functions of the modeling procedure. Here we list
the settings that we use for our standard run. We concentrate on the
settings that are new to the discrete implementation. All other
settings that are needed to fit a \sch model (e.g., the resolution and
limits of the polar grids used to compute the gravitational potential
$\Psi$; the required numerical accuracies in the fitting of the mass
in the meridional plane and/or the projected plane of the sky; etc.)
are identical to previous implementations of the code, so for those we
refer to \citet{vdm98} and \citet{cre99}.

At the heart of the \sch method lies the generation of a comprehensive
library of orbits that should be representative of all types of orbits
possible in the gravitational potential under study. This is achieved
by adequately sampling the ranges of values that the three integrals
of motion $(E,L_z,I_3)$ can acquire, each set of values uniquely
determining one possible orbit. In this work we build models using two
libraries that only differ in their size. Most of our runs consist of
the generation of initial conditions and libraries with
$20\times14\times7 = 1960$ orbits, obtained by sampling the available
integral space with 20 energies $E$, 14 angular momenta $L_z$ (7
positive and 7 negative), and 7 third integrals $I_3$. In order to
study the dependency of the results on the size of the orbit library,
we also compute \sch models using a much larger orbit library, with
$40\times28\times14 = 15680$ combinations of $(E,L_z,I_3)$.

The energy $E$ is sampled via the corresponding radius $R_c$ of the
circular orbit of that energy (that with maximum angular momentum) in
the equatorial plane $(z=0)$. This radius is logarithmically sampled
from a minimum value that we choose to be much smaller than the
spatial resolution of the data, to a maximum value set much beyond the
point at which most of the mass of the input distribution is actually
encompassed. Since totally unconstrained by the data, therefore, the
few first and last energy bins will mostly be of no interest (i.e., no
mass gets assigned to them in the process of optimization). The
vertical component of the angular momentum, $L_z$, is linearly sampled
using the variable $\eta=L_z/L_{\rm max}$, where $\eta \in\,\,(0,1)$
and $L_{\rm max}$ is the angular momentum of the circular orbit with
energy $E$. While orbits with both positive and negative $L_z$ are
included in the library, the latter need not be individually
integrated because they are simply obtained by reversing the velocity
vector at each point along the orbit. The third integral $I_3$ is
sampled via an angle $w \in\,\,(0,w_{\rm th})$, where $w_{\rm th}$
determines the position at which the ``thin tube'' orbit at the given
$(E,L_z)$ touches its zero-velocity curve (defined by the equation $E
= \Psi_{\rm eff}$, where $\Psi_{\rm eff} = \Psi -
\frac{1}{2}L_z^2/R^2$ is the effective gravitational potential; see
\citealt{vdm98} for a detailed presentation). Finally, in order to
help alleviate the discrete nature of the numerical orbit library, some
extra radial smoothing of the orbits is performed by randomly
generating a small variation to the energy and computing and storing
the contribution to the probabilities from the ``new'' orbit with
integrals $(E+\delta E,L_z,I_3)$. It is possible to implement similar
smoothing in $L_z$ and $I_3$ as well (e.g., \citealt{kra05,cap06}),
but we leave this for a future version of our code. This energy
smoothing is repeated, at each timestep, for 7 random $\delta E$
values. Azimuthal averaging is also performed by randomly drawing 7
$\phi$ values at each timestep.

Smoothing in phase-space is accomplished with the use of apertures
(\S\,\ref{sec:pij}). The size of the (squared-shaped) spatial
apertures are defined as a fraction of $R$, the distance to the center
of the stellar system in the projected plane, and we set this fraction
to 10\%. In velocity space, and for most practical applications, the
measurement errors $e_{ik}$ themselves will provide sufficient
``natural'' smoothing for numerical purposes. Thus, we set the factors
$\xi_k$ in equation \ref{eq.weights} to zero for all our tests (see
also discussion in \S\,\ref{sec:pij}). In practice, the optimal value
of $\xi_k$ will depend on the characteristics of the data
(particularly the size of the velocity errors) as well as the stellar
system under study. When dealing with actual data, therefore, at least
a few different values should be tried in order to explore their
impact on the results. Additionally, the extra smoothing provided by
$\xi_k$ can also be useful to explore the validity of the quoted
errors in any given application.

The uncertainties $e_{ik}$ in the LOS velocities and/or proper motions
are in practice determined by the details of the observations and,
since obtained by different techniques (spectroscopy versus
astrometry), are of different size in general. Furthermore, the
uncertainties in the velocities tangential to the plane of the sky are
affected by the uncertainty in the distance to the stellar system
under study. Here, however, since we deal with simulated data, we
assume kinematical data of nowadays typical good quality, and simply
set all these errors to a moderate and arbitrary value of $e_{ik} =
7.1$\kms.

The large majority of our tests were done on the simulated datasets as
described in \S\,\ref{sec:data} {\it without} the addition of
simulated observational errors (i.e., random Gaussian deviates with
dispersion $e_{ik}$). This simplification was made early on in our
project, based on the fact that the velocity errors should not matter
much as long they are much smaller than the average one-dimensional
velocity dispersion of the system under study. However, we realized
later that this does induce a slight bias in our estimated
mass-to-light ratios.  Our typical simulated datasets have dispersions
of 48.4\kms and 46.3\kms, for the 55is and 90is cases,
respectively. Therefore, by not adding any random velocity errors, the
one-dimensional velocity dispersion of the pseudo-data that we
actually analyzed is too small by a factor $h =
(1+(e_{ik}/\sigma)^2)^{1/2}$. As a consequence of the virial theorem,
it follows that we should expect to infer a mass-to-light ratio that
is too small by a factor of $h^2$, corresponding to about 2.2\% and
2.4\% for the 55is and 90is datasets, respectively. Instead of
rerunning all our calculations, which would have been computationally
expensive, we therefore simply corrected for this bias {\it post
facto}. So when studying the recovery of the mass-to-light ratio in
\S\,\ref{sec:getML} and \S\,\ref{sec:grids}, instead of comparing the
inferred values to the value $\Upsilon_0$ of the input model, we
compare to the slightly smaller $\Upsilon_0^* = \Upsilon_0/h^2$. This
quantity is $\Upsilon_0^* = 2.451$ for $i=90^{\circ}$ and
$\Upsilon_0^* = 2.498$ for $i=55^{\circ}$.

The sizes of currently existing kinematic datasets of discrete nature
range from a few hundred datapoints (red giants in Local Group dwarf
galaxies, planetary nebulae in the outskirts of giant ellipticals) to
a few thousands (stars in Galactic globular clusters, systems of
globular clusters around giant ellipticals). For our standard tests we
adopt datasets with 1000 kinematic observational points, although we
also study the consequences of studying datasets with sizes ranging
from 100 to 2000 datapoints. In these tests, the small-$N$ datasets
are subsets of the largest dataset ($N = 2000$), which means that
there will be some correlation between the results of experiments done
as a function of the number of available observations. This approach,
we note, is of no substantial difference than having all the datasets
of different $N$ but within the same simulation to be completely
disjoint. The progression with $N$ should still follow the expected
$N^{-1/2}$ statistical-convergence behavior (see
Fig.~\ref{fig:errors_ML} and \S\,\ref{sec:getML}). The generation of
one of our smaller $20\times14\times7$ orbit libraries, simultaneously
storing discrete probabilities for a set of 1000 observational points
with both LOS velocities and proper motions, takes 2.5 hours on a 3.6
GHz, Pentium 4, 64-bit CPU with 2 Gb memory. An additional 0.5 hours
are needed to find the maximum likelihood fit to the data. In
practice, these steps must be iterated over a grid of gravitational
potential parameterizations.

%.... enforce meridional plane dispersion constraints?: NO ....
%
%.... fractional error in meridional plane mass: 0.05 ....
%
%.... fractional error in vel merid disp constraints: 0.20 ....
%
%.... settings on projected cubes used to store info ....

\subsection{Recovery of the Distribution Function}
\label{sec:getDF}

In order to determine whether the best-fitting solution obtained by
the discrete \sch code actually resembles the properties of the input
data, we start by making detailed comparisons between the input and
fitted DFs. To do this, we feed the code with the correct inclination
and mass-to-light ratio \ml\, used to generate the input datasets, and
compare the fitted mass weights $\zeta_{\rm fit}$ to those
corresponding to the input data, $\zeta_{\rm in}(E,L_z,I_3)$,
approximated using equation (\ref{eq:zetain}). We use datasets with
1000 LOS velocities and proper motions, and present results for both
the small and big orbit libraries detailed in \S\,\ref{sec:settings}.

The comparison is best achieved via the analysis of corresponding one-
and two-dimensional projections of the cubes of mass weights
$\zeta_{\rm fit}(E,L_z,I_3)$ and $\zeta_{\rm in}(E,L_z,I_3)$, obtained
by integrating over two and one of the integrals of motion,
respectively (Figs. \ref{fig:1Dplots} to \ref{fig:2Dplot55is}).  Also,
we make comparisons of two-dimensional $L_z-I_3$ slices of both cubes
at selected values of the energy (Fig. \ref{fig:Ebins55is}). For al
of these projections we quantify the agreement between fits and input
data by computing the RMS and median absolute deviation of the
quantity $(\zeta_{\rm fit}-\zeta_{\rm in})/\zeta_{\rm in}$, i.e., the
difference between fit and input mass weights normalized by the input
mass weights. These statistics are listed for \sch models run on all
our input datasets in Table 1. Since the RMS can be biased
disproportionately by a small number of large outliers, in our
discussion below we use preferentially the median absolute residual.

Figure \ref{fig:1Dplots} shows, for the 55is case, the integrated mass
weights as a function of each of the three integrals of motion, for
both the input dataset and the discrete \sch fit. Inside the region
actually constrained by kinematic data (containing 99.83\% of the
total mass), the mean absolute deviations between the fitted and input
distributions of mass weights are 3\%, 16\%, and 18\%, for the
integrated distributions as a function of $E$, $L_{\rm z}$, and $I_3$,
respectively. As listed in Table 1, similar numbers are obtained for
the other 5 simulated datasets, with the agreement between both
distributions as a function of energy always better than 5\%. As a
function of $L_z$, the largest disagreement actually corresponds to
the one shown in Figure \ref{fig:1Dplots}, the 55is case. It goes down
to 7\% for our case of closest agreement, the case labeled 55ns. The
net rotation inherent to the 55is dataset (reflected in the middle
panel by all the mass weights with positive $L_z$ being larger than
those with negative $L_z$) is clearly reproduced by the \sch fit. As a
function of the third integral, the median absolute deviation varies
from 16\% for the 55ns case to up to 25\% for the 90ns case. Note
that, since we are showing orbital mass weights instead of the actual
DF, the $I_3$ distributions are not expected to be constant over
$I_3$, even though the input DF underlying all simulated datasets is
of the form $f(E,L_z)$.

Next, integrating only over $I_3$, we show in Figures
\ref{fig:2Dplot55ns} and \ref{fig:2Dplot55is} the agreement between
the fitted and input sets of mass weights as a function of $E$ and
$L_z$, for the 55ns and 55is cases, respectively. The upper panels of
these figures show the results of the \sch fit ($\zeta_{\rm fit}$) and
the lower panels the original input distributions ($\zeta_{\rm
in}$). The left-hand panels show the results for a $(E,L_z,I_3)$
library of $40\times28\times14$ orbits, 8 times larger (i.e., finer)
than that of the right-hand panels, which correspond to our standard
case of $20\times14\times7$ orbits.  Only the energy range constrained
by the respective sets of data is shown. Black corresponds to zero
weight, and the white (brightest) color in each pair of panels (fit
and model, or upper and lower) has been assigned to the maximum
orbital weight among the two panels, so that the comparison between
fits and models is made using the same color scale.
%(WILL HAVE TO CHANGE THIS AND SET WHITE TO
%THE MAXIMUM AMONG ALL PANELS ... STILL TO DO).

Both Figures \ref{fig:2Dplot55ns} and \ref{fig:2Dplot55is} show that
the main features of the input $E-L_z$ distributions of mass weights
are well reproduced by the 3-integral \sch fits. In particular, the
mean streaming properties of both datasets are satisfactorily
recovered.  In Figure \ref{fig:2Dplot55ns}, the two prominent
phase-space blobs occupying symmetrical locations on the negative and
positive sides of the $L_z$-axis correspond well with the non-rotating
overall nature of the 55ns dataset. Moreover, this is recovered by
both the models with standard and large orbit libraries (right- versus
left-hand panels). Similarly, in Figure \ref{fig:2Dplot55is}, the
single phase-space blob at positive $L_z$ with a pronounced elongation
towards negative $L_z$ (in light blue and blue), indicative of the
rotating nature of the 55is case, is reproduced by the \sch fit as
well. The median absolute deviations between the fitted and input
$E-L_z$ distributions, always restricted to the energy range
constrained by the data, are 14\% and 19\% for the 55ns and 55is
cases, respectively (Table 1).

In Figure \ref{fig:Ebins55is} we show the 3-dimensional distributions
of mass weights of our 55is case in the form of a series of $L_z-I_3$
planes at different energies. Here again, the upper panels show the
results of the discrete \sch fit ($\zeta_{\rm fit}$), the lower panels
the distribution of mass weights corresponding to the input data
($\zeta_{\rm in}$), and the color scale is set up in the same way as
in the $E-L_z$ figures. As the energy $E$ is sampled via the radius
$R_c$ of the circular orbit (its value in arcmin indicated at the top
of each pair of panels), this series of planes shows the variation of
the $L_z-I_3$ distribution with increasing distance from the center of
the galaxy. The fraction of the total mass at each energy slice is
given as a percentage at the bottom of each panel.

The bottom panels of Figure \ref{fig:Ebins55is} indicate that, in the
inner regions (inside 0.2 arcmin), most of the mass in the 55is
dataset is concentrated in orbits with $L_z$ near zero. The
corresponding upper panels show that the \sch fit recovers this $L_z
\approx 0$ component, but it distributes more weight than the input
model into orbits with positive $L_z$. These inner regions,
nevertheless, have a relatively low mass content in comparison with
regions at larger radii. As the radius increases, the $L_z \approx 0$
region of phase-space gets progressively depleted of stars in favor of
orbits with high $L_z$. This transition is reasonably well reproduced
by the \sch solution, and the agreement between fit and input data
becomes better at large radii, at which point most of the mass at each
energy is concentrated in orbits of high $L_z$.

Note also that a common characteristic of Figures
\ref{fig:2Dplot55ns}, \ref{fig:2Dplot55is}, and \ref{fig:Ebins55is} is
that \sch fits typically present mass weight distributions that appear
broader (more extended) and less peaked than the corresponding
distributions displayed by the pseudo-data. The effect is most obvious
among the right-most panels of Figure \ref{fig:Ebins55is}, where one
can see that the $L_z-I_3$ mass-weight distributions of the input data
(lower panels) have higher peaks and overall sharper features than the
corresponding fitted distributions (upper panels). This is an expected
effect and is due to the combined smoothing of the fitted distribution
introduced both by the (necessary) use of velocity apertures for the
computation of likelihoods (see \S\,\ref{sec:pij}), and by the
regularization constraints imposed in order to enforce smoothness in
phase space. While the first smoothing is particular to our discrete
implementation, the second is a well-known procedure common to most
\sch codes. Models without regularization tend to be unrealistically
noisy \citep{vdm98} and unreliable for parameter estimation
\citep{cre04}. Thus, although we choose to plot the input distribution
of mass weights as they actually are, the most fair of comparisons
would be one in which the \sch fit is compared with a smoothed version
of the original mass weight distribution describing the input data. We
explored this by convolving the input distribution of mass weights
with a (circular) Gaussian kernel, and then computing the same
statistics shown in Table 1 (but this time using the smoothed version
of the input distribution) for different widths of the Gaussian
kernel. We have verified that indeed it is possible to find a kernel
width for which the agreement between fit and input data is best,
improving both the RMS and mean absolute deviations of Table 1 by
factors between 1.2 and 1.5. Finally, we also note that the comparison
in Figure \ref{fig:Ebins55is} might be affected by the accuracy of the
approximation in equation (\ref{eq:zetain}), which means that the
values in Table 1 are actually upper limits to the true accuracy of
the \sch fits.

From these tests we conclude that our discrete \sch code can
successfully recover the original DF inside the region constrained by
the kinematic data, at least for the case in which the inclination and
mass-to-light ratio are assumed known.

\subsection{Recovering the mass-to-light ratio}
\label{sec:getML}

For a large range of potential applications of a \sch code, such as
investigating dark matter halos in galaxies, the most important
property that one is interested in measuring with confidence is the
mass-to-light ratio. In the present tests, this quantity is a scalar,
\ml, although in more general applications it could be a function of
radius. In this section we study in detail the capacity of our code to
infer the correct \ml\, when the inclination of the system is assumed
known. Tests were performed for a number of input datasets in order to
investigate the dependence of the results on key observational
variables such as the number of kinematic measurements and the type of
kinematic constraints available (i.e., only-LOS velocities, only
proper motions, as well as both LOS velocities and proper
motions). All models in this section were computed using our small
orbit library, with $20\times14\times7$ combinations of $(E,L_z,I_3)$.
The results of these experiments are summarized in Figures
\ref{fig:MLparabN}\,-\,\ref{fig:errors_ML}.

For the 90is and 55is cases and using full 3-dimensional velocity
information, Figure \ref{fig:MLparabN} shows the $\Delta \chi^2$
parabolae obtained when applying the discrete \sch code with a number
of \ml\, values, distributed around the correct one ($\Upsilon_0^*$),
for datasets of varying sizes. The quantity
$(\Upsilon/\Upsilon_0^*)^{1/2}$ along the ordinate denotes the
velocity scaling; $(\Upsilon/\Upsilon_0^*)^{1/2} = 1$ corresponding to
a \sch model with the input value $\Upsilon_0^*$, defined in
\S\,\ref{sec:settings}. The zero point of the vertical axis (both in
Figures \ref{fig:MLparabN} and \ref{fig:MLparab2}) is arbitrary, but
the difference $\Delta\chi^2$ between points on the same curve has its
usual statistical meaning, and indeed we compute the (random)
uncertainties on the determination of \ml\, directly from them.

Figure \ref{fig:MLparabN} shows that the difference $\Delta\chi^2$
between points on the same curve becomes larger (the parabolae become
narrower) as the number of available kinematic measurements increases.
The determination of the best-fit \ml\, also depends on the type of
available kinematic measurements. This is illustrated in Figure
\ref{fig:MLparab2}, where we plot the $\Delta \chi^2$ parabolae
obtained when considering only LOS velocities, only proper-motions, or
the full 3-dimensional velocity information. All cases are for the
55is dataset with 1000 kinematic measurements. In this case, the
$\Delta\chi^2$ parabolae become narrower as the number of available
velocity components increases.

Furthermore, the statistical errors are generally smaller for larger
datasets, as well as when more velocity components are available. This
is shown in Figure \ref{fig:errors_ML}, where we plot the behavior of
the best-fit \ml\, and its uncertainties as a function of $\log(N)$,
where $N$ is the number of datapoints. The uncertainties
$\Delta\Upsilon$ displayed in the upper panel of Figure
\ref{fig:errors_ML} represent the $1\sigma$ intervals around the
minimum of the parabolae in Figures \ref{fig:MLparabN} and
\ref{fig:MLparab2}, and are defined as half the distance between the
points on the curve where $\Delta \chi^2 = 1$ with respect to the
minimum. The statistical errors scale roughly as $N^{-1/2}$ over an
interval of 1.3 dex in $\log N$. Also, the errors in the best-fit
\ml\, associated to datasets with only proper-motions (triangles) are
smaller than those associated to only-LOS datasets (open circles) for
any value of $N$.  In other words, our discrete \sch code satisfies
the fundamental statistical expectation that it should become easier
for the method to distinguish between models with different \ml\, when
the amount of observational information is larger. In the case of
datasets with the full 3-d velocity information, the 55is
uncertainties do not quite seem to follow the $N^{-1/2}$ behavior
expected from statistics. We attribute this to our tests having
reached a fundamental floor due to the discrete nature of the models,
a limit that can not be overcome by increasing the number $N$ of
available measurements. This can cause an apparent flattening with
respect to the regular $N^{-1/2}$ behavior at large $N$.

To test the robustness of the errors estimated as above, we performed
the following exercise. Selecting 10 different (disjoint) realizations
of the N-body data (for the 55is case with 1000 measurements of only
line-of-sight velocities, the case most often found in practice), we
repeated the exercise of Figures \ref{fig:MLparabN} and
\ref{fig:MLparab2} and computed discrete \sch models for a set of
different \ml\, values distributed around the correct input one. This
was done using our small orbit library. We obtained an average
best-fit \ml\, of 2.46 (less than $1\sigma$ away from the input value,
$\Upsilon_0^*=2.498$), with an RMS of 0.074 (corresponding to about
3\%). When computing the statistical uncertainties using the $\Delta
\chi^2$ parabolae as described above, the average $1\sigma$ error in
the best-fit \ml\, of the set of experiments turns out to be 0.204,
equivalent to a fractional error of 8\%. This is a factor of 2.5
larger than the scatter in the results from multiple independent
realizations of the pseudo-data. This gap is smaller when additional
information about the individual kinematics of the tracers is
available. Indeed, repeating the above exercise for the same datasets
but now using two-dimensional proper-motions instead of only
line-of-sight velocities, the average error in \ml\, computed from our
$\Delta \chi^2$ parabolae is 0.112, a factor of 1.7 larger than the
scatter of the best-fit values, which was 0.067. Therefore, we
conclude that our error estimation using $\Delta\chi^2$ is
conservative.

Despite the smaller statistical errors for the case with
proper-motions alone, the bottom panel of Figure \ref{fig:errors_ML}
indicates that the best-fit \ml\, is closer to the real value,
$\Upsilon_0^*$, for the case with only-LOS velocities. While the
best-fit \ml\, from datasets with only LOS velocities are well within
$1\sigma$ of $\Upsilon_0^*$ for any $N$, this is not the case for the
datasets with only proper-motions, with best-fit \ml\, values that are
$2-4\sigma$ away from $\Upsilon_0^*$. Still, the formal best-fit \ml\,
for the case of full 3-d velocities (thick squares) is on average
within $2\sigma$ of the real value, $\Upsilon_0^*$, corresponding to
better than $\sim 6$\% accuracy. One contribution to the small
systematic bias in \ml\, may come from the fundamental nature of
inverse problems in general (of which \sch modeling is an example),
namely, that there may not necessarily be a unique solution: it may be
possible to change the mass profile and the DF without appreciably
changing the model predictions. If such is the case and there are
multiple solutions, we do not necessarily expect a flat $\Delta
\chi^2$ profile (i.e., with a number of equally acceptable solutions
containing the correct one), most likely because of numerical noise
and discretization effects. While we cannot rule this out, our results
do show that this probably does not affect the recovered mass-to-light
ratio at more than the $\sim 10$\% level (based on Figure
\ref{fig:errors_ML}, built with models using our smaller orbit
library). Unless superb data are available, random uncertainties are
likely larger than such systematic errors. Currently, the only
exception to this are some Galactic globular clusters, for which
thousands of proper motions are being measured. However, such systems
are often closer to spherical than galaxies, and hence one expects any
theoretical degeneracies to be smaller. Alternatively, numerical noise
in the orbit library may be the cause of this systematic bias in \ml\,
seen in the bottom panel of Figure \ref{fig:errors_ML}.  Numerical
noise may be reduced in part by the use of larger orbit
libraries. Indeed, we show in \S\,\ref{sec:grids} below that a
substantially larger orbit library tends to produce more accurate
results overall.

The likelihood ratio statistic $\Delta \chi^2$ in
Figures~\ref{fig:MLparabN} and \ref{fig:MLparab2} allows us to find
the best-fit model parameters and their confidence intervals.
However, it does not shed light on the question whether the best-fit
model is actually statistically consistent with the data. The
likelihood $\ln L$ of the best-fit model also cannot be used for this
purpose. There is no theorem of mathematics that states what the value
of $\ln L$ should be for a statistically acceptable model, given that
the underlying velocity distributions from which the particles are
drawn are not known a priori (and are not generally Gaussian).
Nonetheless, many other statistics can be defined to address this
issue once the best-fit model has been found. For example, the
velocity moments of the best-fit model can be calculated (as a
function of position on the sky), and statistics can be defined that
assess whether these moments are consistent with the observed data.
Alternatively, one can draw random realizations of the data from the
best-fit model and use a Kolmogorov-Smirnov test to assess whether the
data and the realization are consistent with being drawn from the same
underlying distribution. We have explored a subset of these approaches
and these suggested that the best-fit models are indeed statistically
consistent with the pseudo-data they were designed to fit.

\subsection{Recovering the inclination and M/L}
\label{sec:grids}

In general neither the mass-to-light ratio nor the inclination of a
stellar system under study are known in advance, and thus one has to
explore models with several combinations of both parameters in a
search for those values that provide the best fit to the data. In this
section we present and discuss the results of running the discrete
\sch code on grids of $(i,\Upsilon)$ values to study whether the
correct input combination is recovered. As in \S\,\ref{sec:getML}, we
perform tests on datasets with different types of kinematic
constraints (LOS velocities and/or proper motions).

The results of tests are presented in Figures \ref{fig:55is_LOS_MU}
and \ref{fig:grid_55_90}. They show $\Delta\chi^2$ contours that
result when computing discrete \sch models on a grid of $(i,\Upsilon)$
values, including the correct input combination, for a variety of
input datasets of the 55is and 90is cases. The goodness-of-fit
parameter $\Delta\chi^2$ shown in these plots is obtained by first
rebinning with a much finer grid the $(i,\Upsilon)$ space explored by
the models actually calculated (indicated by small dots), and then
computing the values on this new grid via interpolation between the
nearest calculated models. We then determine the minimum on the finer
grid (whose location is indicated by the star) and subtract it from
all grid points to obtain the $\Delta\chi^2$ parameter, for which
contours are shown. As in the case of Figures \ref{fig:MLparabN} and
\ref{fig:MLparab2}, the mass-to-light ratio is parameterized by the
dimensionless velocity scaling $v_{\rm
s}=(\Upsilon/\Upsilon_0^*)^{1/2}$, so that the input value corresponds
to $v_{\rm s}=1$.

We start by showing in Figure \ref{fig:55is_LOS_MU} the results of
running grids of models for input datasets composed of only-LOS
velocities and only proper motions, in both cases for the 55is case
with 1000 observational datapoints, and using our small orbit library
with $20\times14\times7$ combinations of $(E,L_z,I_3)$. Overall, and
in agreement with the results of Figure \ref{fig:MLparab2} discussed
in \S\,\ref{sec:getML}, the $\Delta\chi^2$ contours indicate that
proper motions (bottom panel) better constrain the best-fit
$(i,\Upsilon)$ combination than a dataset with only-LOS velocities
(upper panel). The $3\sigma$ uncertainties (thick contours) obtained
from the only-LOS dataset are twice as large than those from the
proper motions alone (31\% and 16\%, respectively). The input
mass-to-light ratio $\Upsilon_0^*$ is adequately recovered by both
datasets (to within the $1\sigma$ confidence region). The best-fit
inclination, however, is offset from the actual input value
$i=55\grad$ for both datasets, although somewhat closer to the correct
value in the case of proper motions only. The $3\sigma$ uncertainties
in the best-fit inclination are $\pm 6\grad$ and $\pm 11\grad$ for the
only proper motions and only LOS cases, respectively.

Difficulties in constraining the inclination using \sch modeling of
stellar kinematics have been encountered in the past. A good recent
example is that of \citet{kra05} who, based on integrated stellar LOS
velocity profiles and ionized gas observations of the E4 galaxy NGC
2974, carried out a study analogous to the present one by constructing
simulated observations of this galaxy, which they feed to their
``continuous'' (as opposed to discrete) \sch code in order to study
the recovery of the input mass-to-light ratio and inclination. They
find that even with artificially perfect input kinematics the
inclination is very poorly constrained. The same conclusion is reached
when attempting to fit the actually observed LOS velocity profiles
with \sch models, so stellar LOS velocity profiles provide weak
constraints on the inclination of this system, a statement they are
confident about because the actual inclination of NGC 2974 is known
from observations of its extended disc of neutral and ionized gas in
rapid rotation.

While one could expect that the availability of proper motion
measurements in addition to LOS velocities would enhance the ability
of the models to obtain useful constraints on the inclination of a
stellar system in general, the reality is that the current
state-of-the-art of \sch modeling does not have a definitive answer on
this issue yet. As recent studies of the kinematics of stars in
globular clusters seem to indicate, the chances of success are highly
dependable on the quality and quantity of available data on the system
under study (compare, for example, the results of \citealt{glenn06}
and \citealt{bos06} regarding the best-fit inclinations of $\omega$
Cen and M15, respectively).

There are at least two factors that may contribute to the difficulty
in recovering the inclination from stellar kinematics: degeneracies
inherent to \sch models, and numerical noise. First, there is no
guarantee that inclinations other than the correct one must fit the
data worse. Indeed, in their modeling of high signal-to-noise
integral-field data of NGC 2974, \citet{kra05} already observe that
the differences between \sch models with different inclinations are
smaller than the differences between the best-fitting model and the
data, which they interpret as indication of a fundamental degeneracy
in the recovery of the inclination with three-integral
models. Numerical noise, on the other hand, is a consequence of \sch
models being in the end only discrete representations of a smooth,
continuous distribution of possible orbits, and it could be argued
that this discreteness might have a more negative effect for high
inclinations. For example, even a simple and smooth circular orbit
presents cusps or discontinuities when viewed close to edge-on. The
turning points of such an orbit may get smoothed out differently for
different inclinations.

The issue of degeneracy, nevertheless, can be avoided in those cases
where the inclination is known to be uniquely determined by the
data. This is the case, e.g., in the situations where the following
conditions are met: (1) the kinematical dataset consists of proper
motion measurements and LOS velocities, (2) the stellar system is
reasonably close to axisymmetric, and (3) there exists an independent
measurement of the distance $D$ to the system. As first used in
practice by \citet{glenn06}, the inclination then follows directly
from the following relationship between the mean LOS velocity (in
units of \kms) and the mean proper motion along the short axis (in
units of mas\,yr$^{-1}$),
\begin{equation}
\label{eq:inclination}
\langle\,v_{z'}\,\rangle = 4.74\,\,D\,\tan i\,\,\langle\,\mu_{y'}\,\rangle, 
\end{equation}
where $D$ is the distance in kpc, and the brackets denote an
integration along the line-of-sight. This relation is true at each
projected position $(x',y')$ in any axisymmetric system, and has been
successfully applied to the Galactic globular clusters $\omega$ Cen
and M15 \citep{glenn06,bos06}.

Here, in order to explore the applicability of this simple
relationship, we take advantage of our a priori knowledge of the
correct inclination for our simulated datasets, and study the
circumstances under which the use of equation (\ref{eq:inclination})
provides an accurate result. Unlike the case of integrated light
measurements (where $\langle\,v_{z'}\,\rangle$ is simply the average
of the LOSVD at any given projected position on the sky), in the
context of discrete datasets neither $\langle\,v_{z'}\,\rangle$ nor
$\langle\,\mu_{y'}\,\rangle$ are quantities that can be rigorously
obtained from the data at any given $(x',y')$. Both quantities may,
nevertheless, be approximated by averaging a number of kinematic
measurements that fall within one or more apertures of a given size
around projected positions $(x',y')$. Following this, we applied
equation (\ref{eq:inclination}) to a series of subsets of our 6
simulated datasets with varying number of kinematic measurements, and
verified that indeed the correct inclination is reproduced provided:
(a) the system is rotating (otherwise, while the relation is still
valid, both averages are nearly zero and hence the inclination is not
really constrained); (b) most of the datapoints are not located close
to the minor axis (where rotation velocities are too small); and (c)
the averages are computed from a sufficiently large number of
kinematical measurements (so that the error in $\tan i$ is not too
large). These are conditions that are certainly fulfilled by datasets
on some Galactic globular clusters, currently the only class of
stellar system for which there are 3-dimensional kinematic information
available. Therefore, in those cases, equation (\ref{eq:inclination})
can be safely applied. The \sch modeling can then concentrate on
recovering the more interesting properties such as the orbital
structure and mass-to-light ratios, which we have shown are
successfully recovered when the inclination is assumed known.

To better understand the problem of numerical noise, we explored the
dependence of the results on the size of the orbit library used to
construct the \sch models. We did this for cases with 1000 datapoints
with complete three-dimensional velocities, so that because of
equation (\ref{eq:inclination}) we know that there is no theoretical
degeneracy in inclination. Figure \ref{fig:grid_55_90} shows the
$\Delta\chi^2$ contours resulting from fits of \sch models using our
standard library of $20\times14\times7$ orbits (upper panels; same
library size as in Figure~\ref{fig:55is_LOS_MU}) in comparison with
fits that use a library 8 times larger, i.e., one with
$40\times28\times14$ orbits (lower panels). We show results for the
55is (left-hand panels) and 90is (right-hand panels) cases.

In all four panels of Figure \ref{fig:grid_55_90}, the best-fit
mass-to-light ratio is always within $1\sigma$ of the input value
$\Upsilon_0^*$, with the exception of the 55is case with the bigger
library (lower left), where they agree at the $2\sigma$ level. The
size of the confidence regions on the mass-to-light ratio does not
change significantly when the orbit library is increased in size.
Therefore, we conclude that libraries of $20\times14\times7$ orbits
are large enough to properly constrain the mass-to-light ratio
(provided that one uses regularization as we do here; see
\citealt{cre04}). This provides further justification for our use of
this library size in Section~\ref{sec:getML}.

The top left panel in Figure \ref{fig:grid_55_90} is directly
comparable to the two panels of Figure \ref{fig:55is_LOS_MU}, but now
with three components of velocities observed, instead of just one or
two, respectively. Consistent with the results in
Figures~\ref{fig:errors_ML} and~\ref{fig:55is_LOS_MU}, we see that the
addition of an extra component of velocity decreases the size of the
confidence regions. More interestingly, a secondary minimum in $\Delta
\chi^2$ appears close to the $(i,\Upsilon)$ values for the correct
input model. This suggests that indeed all three components of
velocity may be necessary to uniquely constrain the inclination of an
axisymmetric stellar system. The bottom left panel shows the effect of
increasing the orbit library size. There is now only a single minimum,
centered at an inclination that agrees with the input value at the
$\sim 2\sigma$ level.

The right panels in Figure \ref{fig:grid_55_90} show the situation for
the 90is case. With the small library (top right), the best-fit
inclination is at $i\approx 70\grad$, substantially far from the input
value. When the orbit library size is increased (lower right), the
best-fit shifts to $i=80\grad$. This is only $10\grad$ from the
correct input value, which may well be acceptable for many realistic
applications. On the other hand, the best fit and the input value are
inconsistent at the many sigma level, which is certainly reason for
some concern. A possible cause for this is that the turning points of
orbits in edge-on systems have very sharp edges in
projection. Therefore, larger grid sizes than we have used may be
necessary to correctly represent them in all the necessary detail.
However, we have not explored this further for two reasons. First,
information on all three velocity components may be necessary to be
able to uniquely constrain the inclination. If that is available, then
use of equation~(\ref{eq:inclination}) will be more accurate and
efficient than use of Schwarzschild modeling. Second, in practice one
is generally much more interested in the mass distribution than in the
inclination. Figure \ref{fig:grid_55_90} shows that the mass-to-light
ratio is correctly recovered, even when the inclination is
systematically biased.

In conclusion, our tests demonstrate that the recovery of the most
important properties of the system (its orbital structure and the
mass-to-light ratio) by our discrete \sch models is robust. Correct
recovery of the inclination appears to be the most complicated aspect
of the modeling. Sufficient observational data must be available and a
large enough orbit library must be used. Our code can then adequately
recover the inclination of sufficiently inclined systems. However, for
edge-on systems there remains a systematic inclination bias of $\sim
10\grad$ that we have been unable to resolve. This is the primary
shortcoming of our new approach that was unearthed by the pseudo-data
tests that we have presented. This may be a generic property of
Schwarzschild codes, since other authors have also reported
difficulties in recovering inclinations. Either way, this is not
believed to be a significant limitation for most potential practical
applications of our code.

\section{Summary and conclusions}
\label{sec:end}

Discrete kinematic datasets, composed of velocities of individual
tracers (e.g., red giants, planetary nebulae, globular clusters,
galaxies, etc.), are routinely being assembled for a variety of
stellar systems of all scales (\S\,\ref{sec.intro}). These include not
only LOS-velocity surveys. High-quality proper-motion databases
already exist for Galactic globular clusters, and future facilities
hold the promise of providing the same for stars in the nearest
galaxies. However, the most sophisticated tools typically being used
in the modeling of these observations were actually developed for the
analysis of kinematic data in the form of LOSVDs, a rather different
type of velocity information than the case of the velocities of
kinematic tracers on a one-by-one basis. As a consequence, the
information content of any particular dataset of a discrete nature is
likely not being fully exploited. We thus have developed a specific
tool for the modeling of discrete datasets, which we have presented in
this paper along with detailed tests of its performance based on the
modeling of simulated data.

The new tool consists of a \sch orbit-superposition code that, adapted
from the implementation of \citet{vdm98}, can handle any number of
(one-, two-, or three-dimensional) velocities of individual kinematic
tracers without relying on any binning of the data. Under the only
assumptions that the system is in steady-state equilibrium (i.e., the
gravitational potential is not changing in time) and may be well
approximated as axisymmetric, the code finds the distribution function
(a function of the three integrals of motion $E$, $L_z$, and $I_3$)
that best reproduces the observations (the velocities of the tracers
as well as the overall light distribution) in a given potential. The
fact that the distribution function is free to have any dependence on
the three integrals of motion allows for a very general description of
the orbital structure, thus avoiding common restrictive assumptions
about the degree of (an)isotropy of the orbits.

Unlike previous implementations of the \sch technique, we cast the
problem of finding the best superposition of orbits using a
probabilistic approach, i.e., by building a likelihood function
representing the probability that the entire set of measurements would
have been observed assuming a particular form for the gravitational
potential (\S\,\ref{sec:logL}). In this case, and in contrast with the
old continuous versions, the dependence of the likelihood function on
the orbital weights is non-linear, and the optimization problem can
not be reduced to a linear matrix equation. Instead, it becomes a
problem of the maximization of a likelihood with respect to the set of
weights associated to all possible combinations of the integrals
$(E,L_z,I_3)$ that comprise the orbit library (\S\,\ref{sec:logL}),
and which accounts for the observed positions and (any-dimensional)
velocities of all particles in the dataset, including their
uncertainties (\S\,\ref{sec:pij}). After extensive testing, a
conjugate gradient algorithm was found to converge satisfactorily to
the correct solution and was adopted for the remaining tests of the
code's overall performance (\S\,\ref{sec.mkfitin}).

In order to assess the reliability of our discrete \sch code, we
applied it to several sets of simulated data, i.e., artificially
generated kinematic observations obtained from a model of an
axisymmetric galaxy of which the orbital structure, mass distribution,
and inclination are known in advance. Pseudo-datasets were generated
from a two-integral phase-space distribution function with varying
degrees of overall rotation, types of velocity information (only-LOS,
only proper motions, and both), total number of particles, and for two
different inclinations on the plane of the sky (\S\,\ref{sec:data}).

Using the various simulated datasets, we studied the recovery of the
input orbital structure or DF, mass-to-light ratio, and
inclination. For the purposes of these tests, we assumed complete
knowledge of the radial profile of the underlying mass distribution
and a mass-to-light ratio that remains constant as a function of
radius. These restrictions are easily (and must be) lifted when
modeling data on real systems, in which case one needs to explore a
range of plausible underlying potentials and allow for variations of
the mass-to-light ratio to properly account for the possibility of
central black holes and dark halos.

\begin{figure}
\vspace*{-1cm}
%\plotone{ELzI3_55is.ps}
\plotone{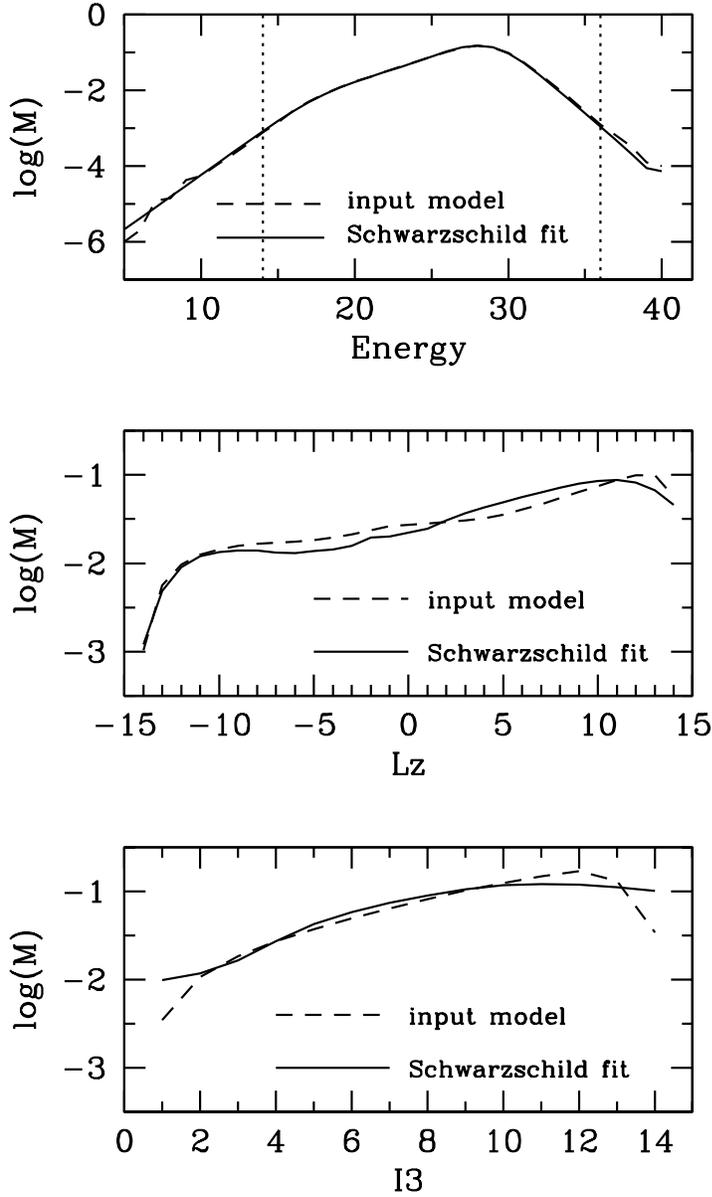}
%{\hspace{-15cm}
%\plotfiddle{ELzI3_55is.ps}{1cm}{0}{400}{500}{-25}{0}}
%\vspace*{-3cm}
\caption{\label{fig:1Dplots} Integrated mass weights as a function of
the three integrals of motion, $E$, $L_z$, and $I_3$, for the 55is
dataset ($i=55^{\circ}$ with intermediate streaming) with 1000
kinematic constraints and full 3-dimensional velocities (both LOS
velocities and proper motions). The \sch fit (solid lines) was
obtained using a library with $40\times28\times14$ orbits (our
``large'' library), and satisfactorily reproduces the mass
distributions associated with the input dataset (dashed lines). The
vertical dotted lines in the upper panel indicate the energy range
constrained by the kinematic data.  The middle panel, with the mass
distribution at positive $L_z$ always higher than that at negative
$L_z$, reflects the net rotation of the 55is dataset. Note that, since
we are showing orbital mass weights instead of the actual distribution
function, the $I_3$ distributions in the bottom panel are not constant
over $I_3$, even though the input distribution function is of the form
$f(E,L_z)$.}
\end{figure}

\begin{figure}
\vspace*{-1cm}
%\plotone{FIGURE_DF.55ns.model_vs_fit.En_Lz_2D.40x14x14_vs_20x7x7.ps}
\plotone{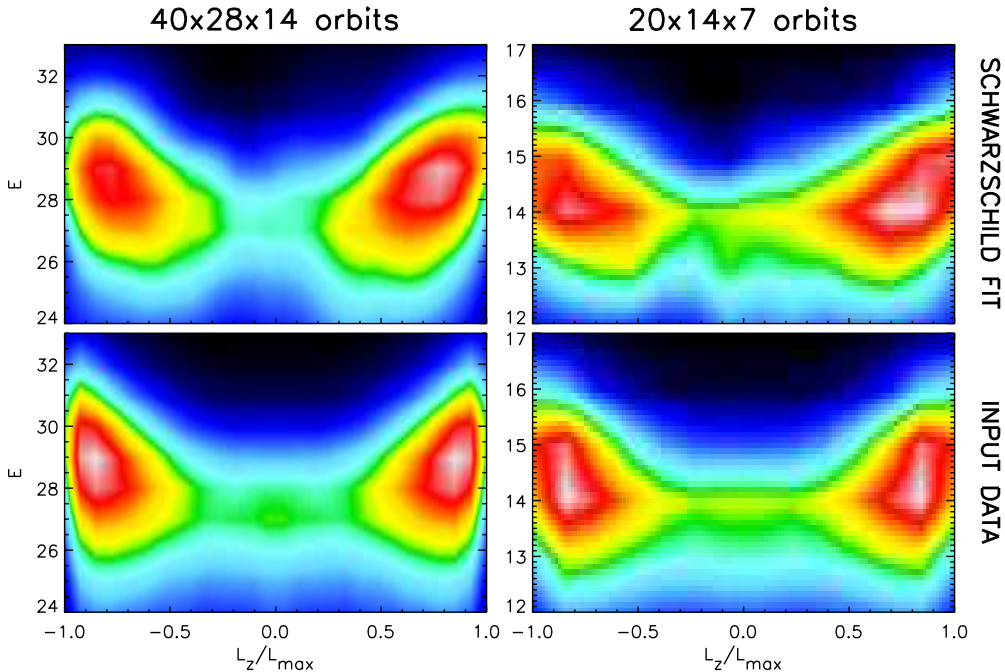}
%{\hspace{-15cm}
%\plotfiddle{FIGURE_DF.55ns.model_vs_fit.En_Lz_2D.1st_version.40x14x14.ps}{1cm}{0}{400}{500}{-25}{0}}
%\vspace*{-3cm}
\caption{\label{fig:2Dplot55ns} Comparison of the input and fitted
distributions of mass weights as a function of energy and $L_z$ for
the 55ns dataset ($i=55^{\circ}$ with no streaming) with 1000 LOS
velocities and proper motions. Only the energy range containing most
of the total mass is shown. Upper panels show the weight distribution
obtained by the \sch fit when the inclination and mass-to-light ratio
are assumed to be known a priori, and the bottom panels show the
weights distribution associated to the simulated input
data. Right-hand panels show the results of the \sch code for an orbit
library with $20\times14\times7$ orbits, while the left-hand panels
show the results for a library 8 times bigger, with
$40\times28\times14$ combinations of $(E,L_z,I_3)$. Black corresponds
to zero weight, and the white (brightest) color in each pair of panels
(fit and model, or upper and lower) has been assigned to the maximum
orbital weight among the two panels, so that the comparison between
fits and models is made using the same color scale. The images in this
and subsequent figures are based on two-dimensional spline curves
fitted to the gridded information. While the two bottom panels
represent the same input data, their visualizations differ due to a
different coarseness in the gridding of integral space.}
\end{figure}

\begin{figure}
\vspace*{-1cm}
%\plotone{FIGURE_DF.55is.model_vs_fit.En_Lz_2D.40x14x14_vs_20x7x7.ps}
\plotone{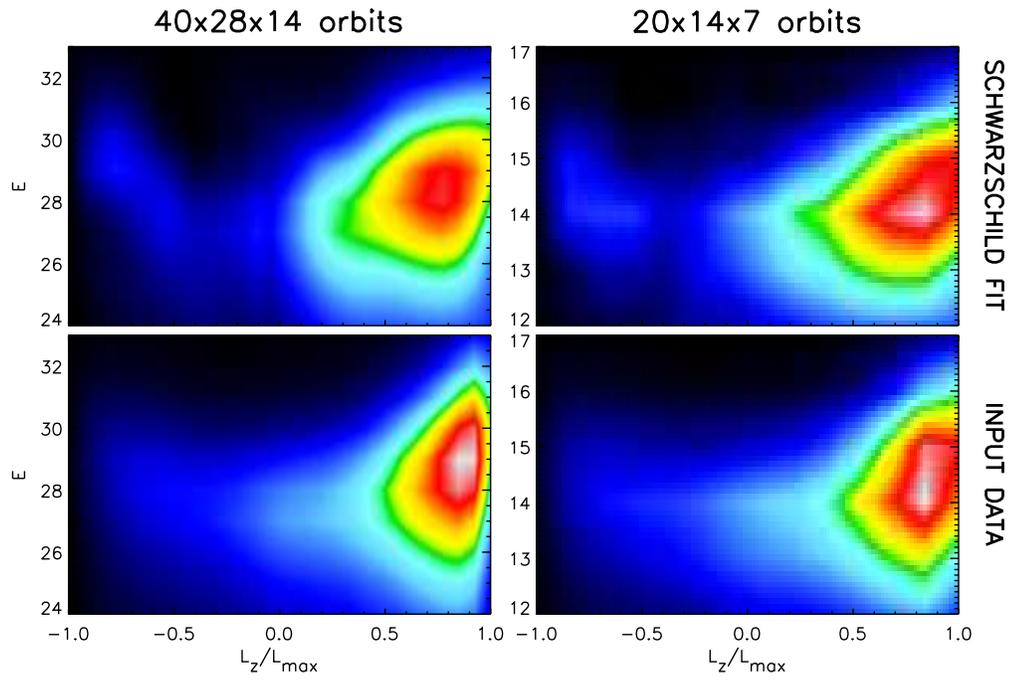}
%{\hspace{-15cm}
%\plotfiddle{FIGURE_DF.55ns.model_vs_fit.En_Lz_2D.1st_version.40x14x14.ps}{1cm}{0}{400}{500}{-25}{0}}
%\vspace*{-3cm}
\caption{\label{fig:2Dplot55is} Same as in Figure \ref{fig:2Dplot55ns} but for
the 55is dataset ($i=55^{\circ}$ with intermediate streaming).
}
\end{figure}

\begin{figure}
\vspace*{-1cm}
%\plotone{FIGURE_DF.55is.model_vs_fit.10_E_bins.40x14x14.ps}
\plotone{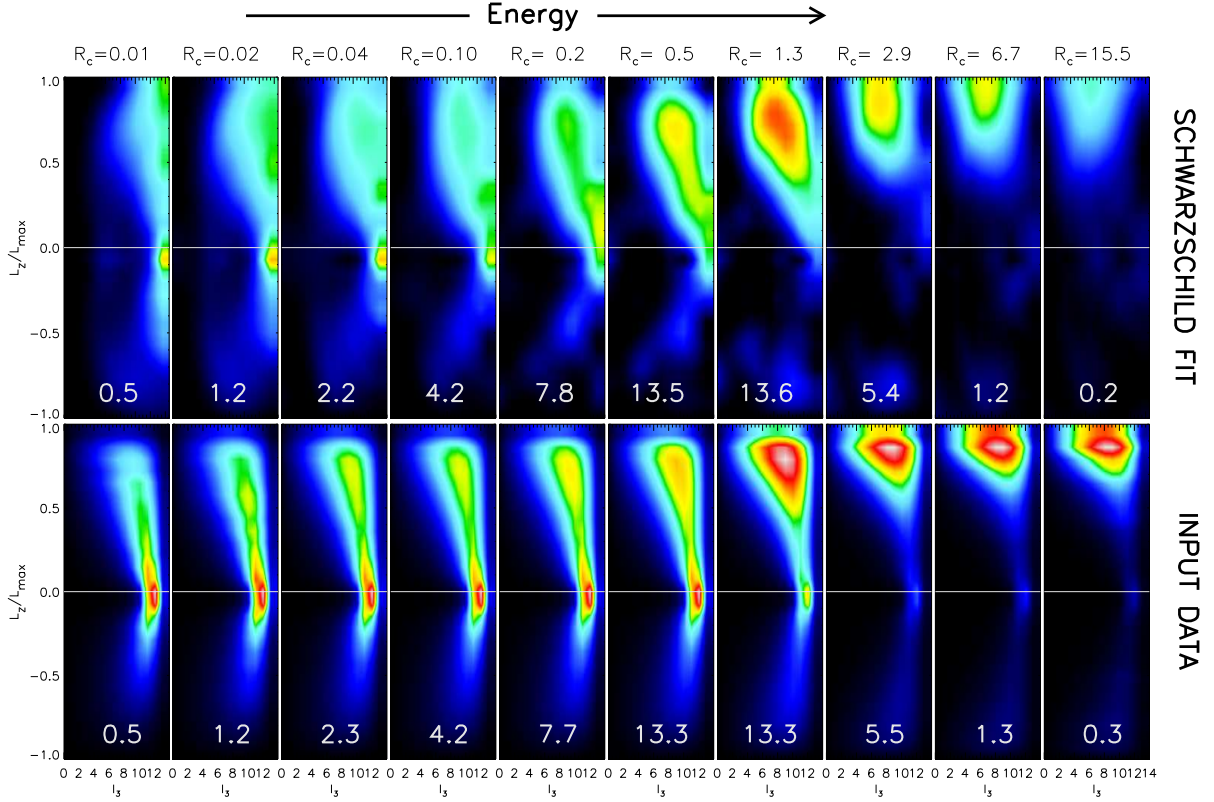}
%{\hspace{-15cm}
%\plotfiddle{FIGURE_DF.55is.model_vs_fit.10_E_bins.40x14x14.ps}{1cm}{0}{400}{500}{-25}{0}}
%\vspace*{-3cm}
\caption{\label{fig:Ebins55is} Input ($\zeta_{\rm in}$; bottom) and
fitted ($\zeta_{\rm fit}$; top) distributions of orbital mass weights
as a function of $L_z$ and $I_3$ at fixed (non-consecutive) values of
energy, for the 55is case with 1000 kinematic measurements with LOS
velocities and proper motions (i.e., the same case as depicted in
Figure \ref{fig:2Dplot55is}). These results correspond to our orbit
library with $40\times28\times14$ combinations of $(E,L_z,I_3)$. From
left to right, the panels show the weight distribution at increasing
distances from the center of the galaxy, as indicated at the top of
each pair of panels by the value $R_c$ (in arcmin) of the circular
orbit at the corresponding energy. The fraction (in \%) of the total
mass contained in each energy slice is indicated at the bottom of each
panel. As in Figures \ref{fig:2Dplot55ns} and \ref{fig:2Dplot55is},
black corresponds to zero weight and the white (brightest) color in
each pair of panels (fit and model, or upper and lower) has been
assigned to the maximum orbital weight among the two panels, so that
the comparison between fits and models is made using the same color
scale. }
\end{figure}

Inside the region constrained by data, we find that the distribution
function (represented by the corresponding distributions of orbital
mass weights) and streaming characteristics of the input datasets are
satisfactorily recovered by the \sch fits when the correct inclination
and mass-to-light ratio are known (Figs. \ref{fig:1Dplots} to
\ref{fig:Ebins55is}). As measured by the mean absolute deviations
between the integrated weight distributions, the agreement between the
fitted and the input orbital weight distributions as a function of
$E$, $L_z$, and $I_3$ is typically of the order of 3\%, 10\%, and
20\%, respectively (the numbers for our worst case being 5\%, 16\%,
and 25\%). When eliminating the dependence on $I_3$, the agreement
between the fitted and input $E-L_z$ distributions is of the order of
15\%, with the net rotational behavior of the input datasets cleanly
recovered (Figs.  \ref{fig:2Dplot55ns} and
\ref{fig:2Dplot55is}). Thus, we conclude that the discrete \sch code
can successfully recover the orbital structure of the system under
study.

\begin{figure}
\vspace*{-1cm}
%\plotone{ML_parabolas.N.eps}
\plotone{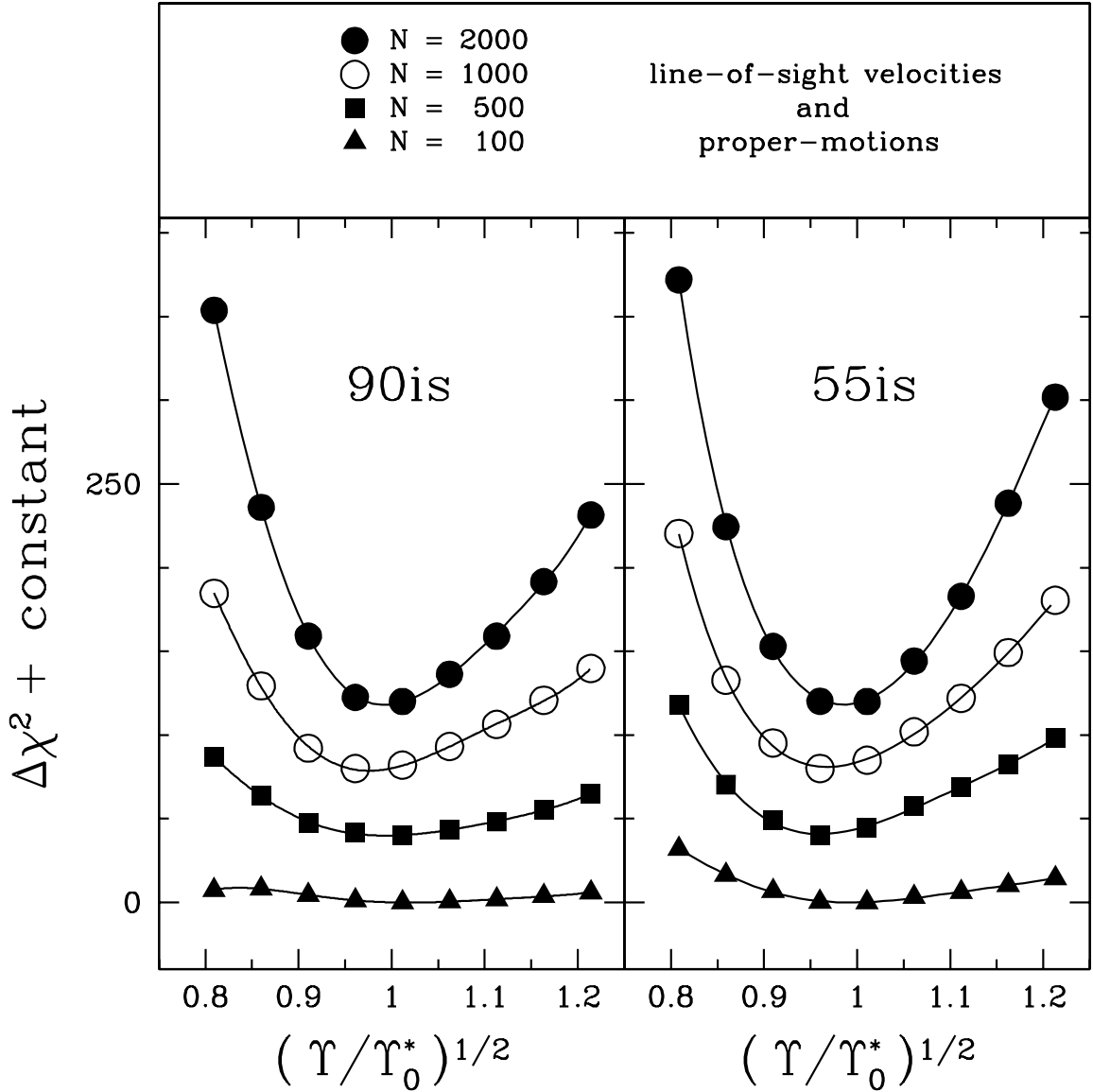}
%{\hspace{-15cm}
%\plotfiddle{ML_parabola.55is_90is.40x14x14.eps}{1cm}{0}{400}{500}{-25}{0}}
%\vspace*{-3cm}
\caption{\label{fig:MLparabN} $\Delta \chi^2-$parabolae that
illustrate the recovery of the input mass-to-light ratio
$\Upsilon_0^*$ as a function of the number of available kinematic
measurements. All input datasets include both LOS velocities and
proper motions, and all \sch models have been computed using our small
orbit library, the one with $20\times14\times7$ combinations of the
$(E,L_z,I_3)$ integrals of motion. For any given input dataset, the
symbols show the $\Delta \chi^2$ obtained by the discrete \sch code on
a number of \ml\, values distributed around the correct one
($\Upsilon_0^*$). The curves connecting the computed models are
polynomial fits of 5th order. When the number of datapoints $N$ is
smaller, the $\Delta \chi^2$ parabola is shallower, and the
statistical uncertainty on the inferred $\Upsilon$ is larger. The
lowest curve is shown at its actual $\Delta \chi^2$.  Each subsequent
curve was offset vertically by a value of 40 for visual clarity.}
\end{figure}

\begin{figure}
\vspace*{-1cm}
%\plotone{ML_parabolas.type.eps}
\plotone{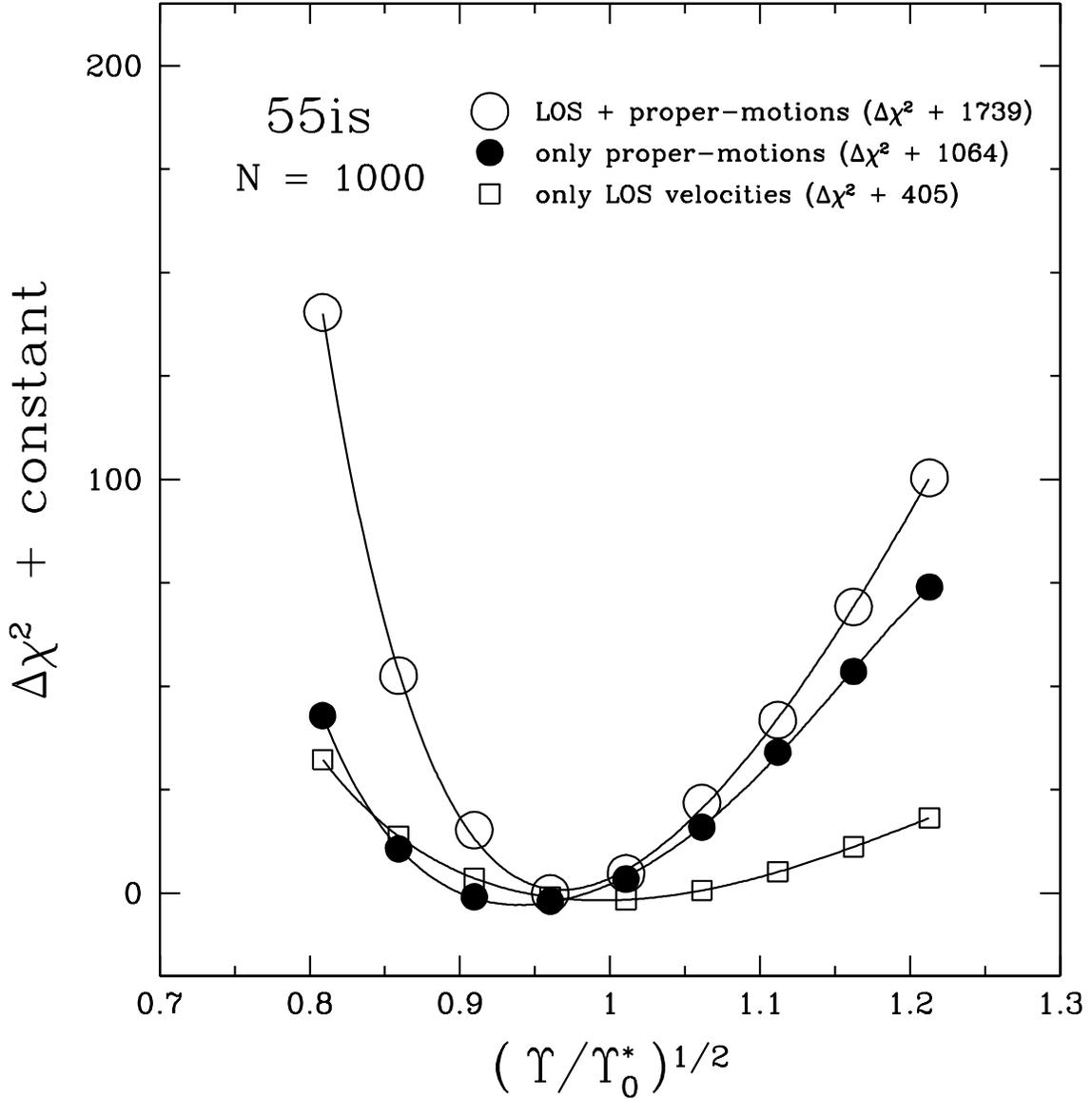}
%{\hspace{-15cm}
%\plotfiddle{ML_parabola.55is_90is.40x14x14.eps}{1cm}{0}{400}{500}{-25}{0}}
%\vspace*{-3cm}
\caption{\label{fig:MLparab2} $\Delta \chi^2-$parabolae illustrating
the recovery of the input mass-to-light ratio $\Upsilon_0^*$ for
datasets with different types of kinematic information. All input
datasets are of the 55is case ($i=55\grad$ with intermediate
streaming) with 1000 measurements. As in Figure \ref{fig:MLparabN},
all \sch models have been computed using our small orbit library, with
$20\times14\times7$ combinations of the $(E,L_z,I_3)$ integrals of
motion.  When fewer velocity components are observed, the $\Delta
\chi^2$ parabola is shallower, and the statistical uncertainty on the
inferred $\Upsilon$ is larger. 
%The lowest curve is shown at its actual
%$\Delta \chi^2$.  Each subsequent curve was offset vertically by a
%value of 50 for visual clarity.
}
\end{figure}

\begin{figure}
\vspace*{-1cm}
%\plotone{errors_ML.eps}
\plotone{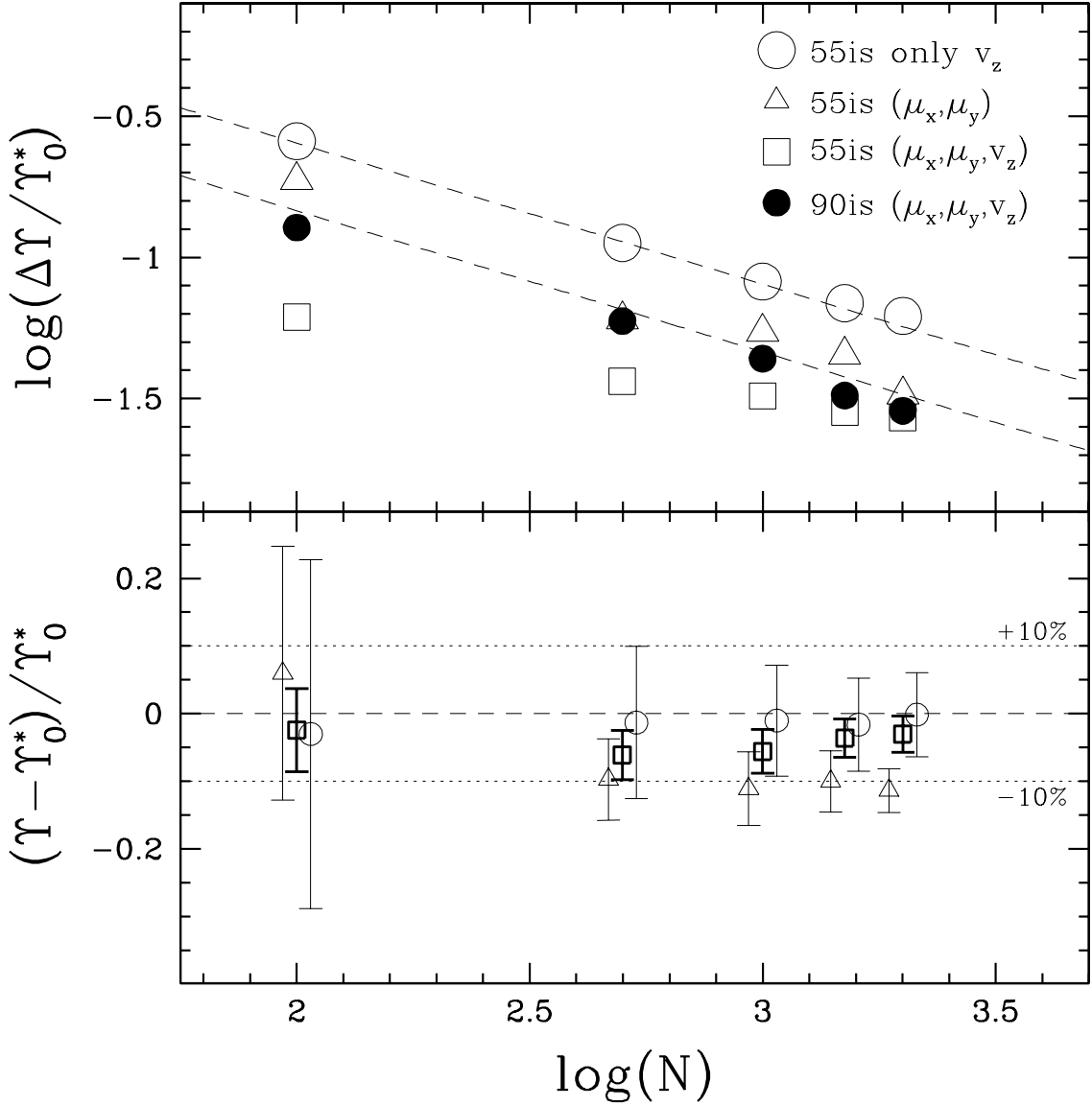}
%{\hspace{-15cm}
%\plotfiddle{ML_parabola.55is_90is.40x14x14.eps}{1cm}{0}{400}{500}{-25}{0}}
%\vspace*{-3cm}
\caption{\label{fig:errors_ML} Uncertainties in the recovery of the
input mass-to-light ratio $\Upsilon_0^*$ as a function of the number
of available kinematic measurements, and for input datasets with
varying types of kinematic information. The upper panel shows the
behavior of the statistical uncertainty in the determination of the
best-fit \ml, i.e., the $1\sigma$ interval around the minimum of the
corresponding parabolae in Figures \ref{fig:MLparabN} and
\ref{fig:MLparab2}. The dashed lines in the upper panel have a slope
of $-1/2$ and serve to demonstrate that the errors given by the \sch
code roughly satisfy the $N^{-1/2}$ scaling expected from number
statistics. The bottom panel shows the difference between the input
mass-to-light ratio ($\Upsilon_0^*$) and the best-fit \ml\, given by
the \sch code (i.e., the minimum of the parabolae of Figures
\ref{fig:MLparabN} and \ref{fig:MLparab2}). The error bars are the
$1\sigma$ errors from the upper panel. All \sch models in this figure
have been computed using our small orbit library, with
$20\times14\times7$ combinations of the $(E,L_z,I_3)$ integrals of
motion. }
\end{figure}

Assuming that the inclination of the system on the plane of the sky is
known, we quantified the recovery of the input mass-to-light ratio as
a function of the size of the input dataset (Fig. \ref{fig:MLparabN})
and of the type of kinematic information available
(Fig. \ref{fig:MLparab2}). We studied both the best-fit value as well
as the uncertainty in its determination
(Fig. \ref{fig:errors_ML}). The statistical expectation of better
results when the amount of observational information is larger (either
regarding the number of datapoints or the number of velocity
components) is clearly reproduced by our discrete \sch models. For the
smallest datasets used in our testing ($N=100$), and regardless of
whether using only-LOS velocities, only proper motions, or both, the
best-fit mass-to-light ratio is within 5-10\% of the input value, with
formal $1\sigma$ uncertainties of the order of 15\%. When increasing
either the number of available measurements or the number of measured
velocity components, the mass-to-light ratio is always recovered to
better than $\sim 10\%$ accuracy, with the corresponding random
($1\sigma$) uncertainties in the range of 5-10\%. The discrete \sch
code, therefore, recovers the mass-to-light ratio of the input
datasets to satisfactory levels of accuracy.

\begin{figure}
\vspace*{0cm}
%\plotone{delta_chisqr.fine_grid.55is_los_mu.eps}
\plotone{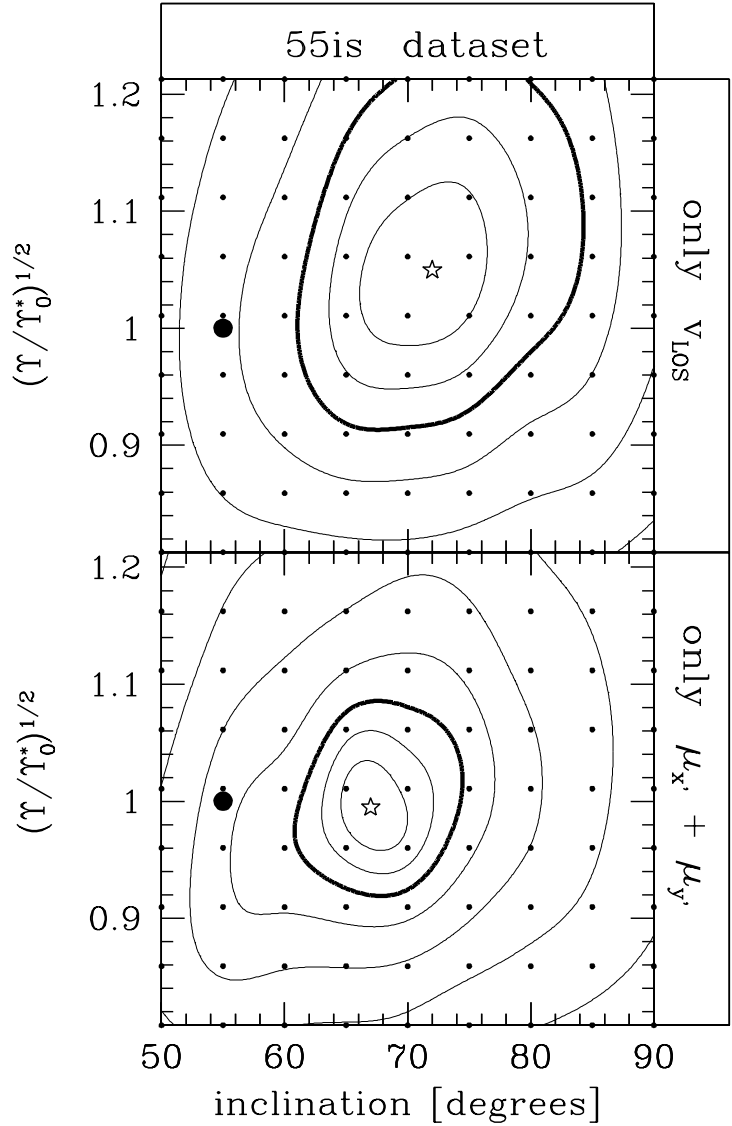}
%{\hspace*{-18cm}
%\plotfiddle{55is.onlyLOS_vs_onlyMU.1000.ps}{1cm}{0}{1.0}{1.0}{0}{0}}
%\vspace*{-3cm}
\caption{\label{fig:55is_LOS_MU} Comparison of discrete \sch models
based on data comprised of purely LOS velocities (upper panel) and
purely proper motions (lower panel), for the 55is dataset with 1000
kinematic measurements and libraries with $20\times14\times7$
orbits. The lines are $\Delta\chi^2$ contours overlaid on grids of
actually computed models (indicated by the small dots) with different
combinations of inclination and mass-to-light ratio \ml. The correct
input model is indicated as a large black dot, and the best-fit model
as a star.  The first three contours are spaced in increments of
$1\sigma$ confidence, with the $3\sigma$ contour (99.7\% confidence
level) highlighted with a thick line. Discrete \sch fits on both
only-LOS velocities and only proper motions satisfactorily recover the
input mass-to-light ratio, but not the input inclination. In terms of
the uncertainties in the best-fit parameters (i.e., the size of the
confidence intervals), proper motions provide tighter constraints than
only-LOS velocities. }
\end{figure}

\begin{figure}
\vspace*{-25mm}
%\plotone{delta_chisqr.fine_grid.library_size.eps}
\plotone{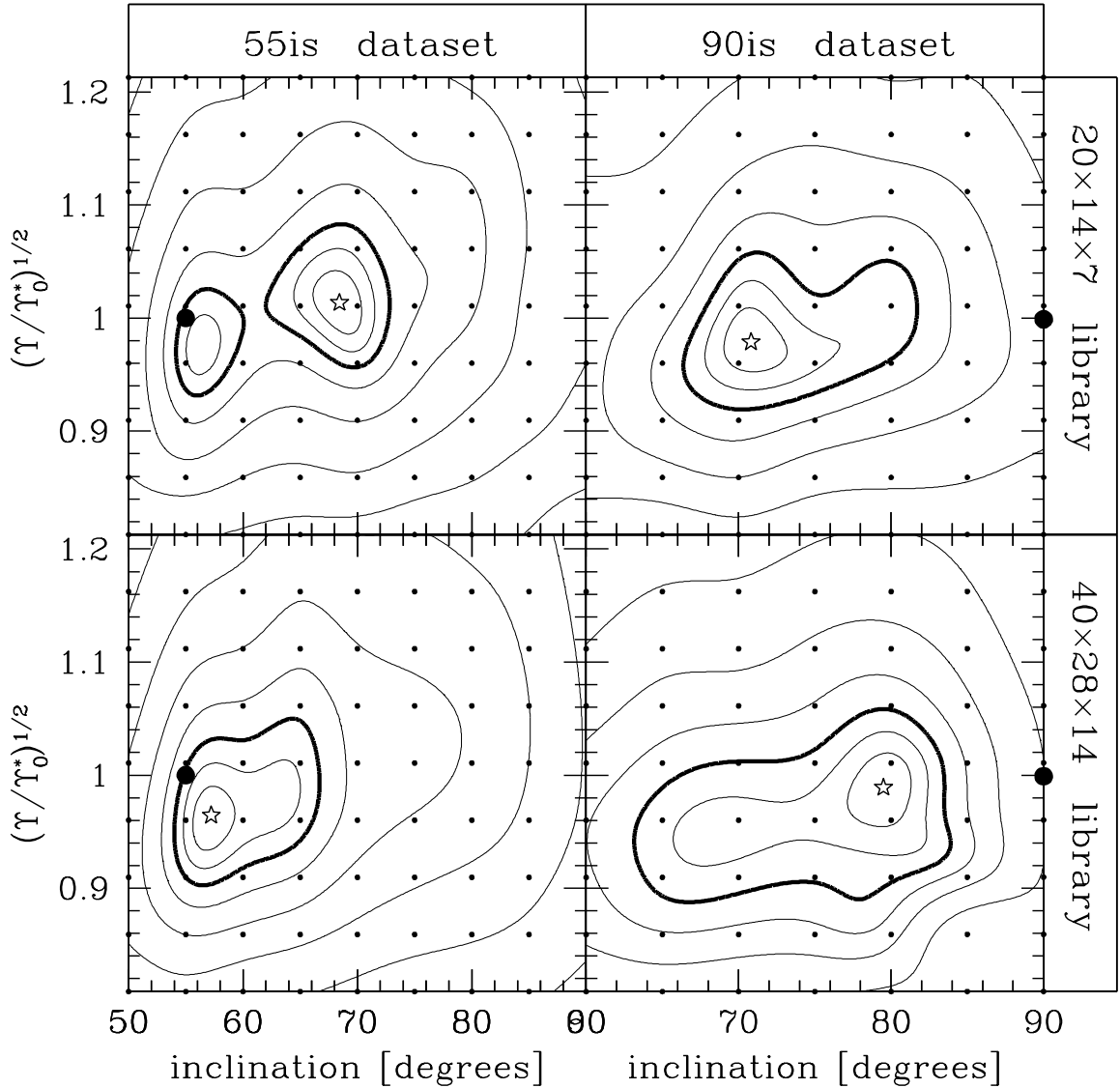}
%{\hspace{-15cm}
%\plotfiddle{}{1cm}{0}{400}{500}{-25}{0}}
%\vspace*{-3cm}
\caption{\label{fig:grid_55_90} Recovery of the input inclination and
mass-to-light ratio, when both assumed unknown, for the 55is and 90is
datasets (left- and right-hand panels, respectively). Shown are the
$\Delta\chi^2$ contours obtained from grids of \sch models constructed
using orbit libraries with different sampling of the available
$(E,L_z,I_3)$ integral space. Upper panels correspond to libraries
with $20\times14\times7$ orbits, while lower panels are based on
libraries with $40\times28\times14$ orbits, i.e., with 8 times finer
sampling.  The input mass-to-light ratio $\Upsilon_0^*$ is
satisfactorily recovered regardless of the number of orbits (in all
cases inside the $2\sigma$ confidence level). In terms of inclination,
the shapes of the contours indicate that there may be two separate
maxima providing similarly good fits to the data. For the smaller
orbit library, the best-fit inclinations converge to the wrong
solution, $i\approx 70\grad$, for both datasets. Nevertheless, the
correct inclination is encompassed by the secondary maximum in the
55is case (upper left), and a clear elongation of the contours towards
higher inclination is seen in the 90is case (upper right). When using
the larger orbit library, however, the best-fit inclination is
$i=57\grad$ for the 55is dataset, and $i=80\grad$ for the 90is
dataset, in better agreement with the true values.  }
\end{figure}

The recovery of both the mass-to-light ratio and inclination when
neither of these quantities are known in advance (as is usually the
case with real observations) was studied using a grid of discrete \sch
models, exploring also the dependence on the type of velocity
components available (Fig. \ref{fig:55is_LOS_MU}). We find that the
mass-to-light ratio was again successfully recovered, but the best-fit
inclination was not identified correctly using small orbit
libraries. We found that this was remedied by better sampling the
available $(E,L_z,I_3)$ integral space using a larger orbit library
(Fig. \ref{fig:grid_55_90}). For our input datasets with $i=55\grad$,
the best-fit inclination obtained by our models with a large orbit
library is $57\grad$, while for input datasets with $i=90\grad$ we
obtain a best-fit model with $i=80\grad$. Given the known difficulty
of \sch models in general for determining the inclination of stellar
systems, and considering the low relative importance of this parameter
compared to other properties such as the orbital structure and the
mass-to-light ratio, we regard this small disagreement for the high
inclination datasets as acceptable.

In summary, we have shown that our new \sch code, designed to
adequately handle modern datasets composed of discrete measurements of
kinematic tracers, doing this without any loss of information due to
data binning or restrictive assumptions on the distribution function,
is able to constrain satisfactorily the orbital structure,
mass-to-light ratio, and inclination of the system under
study. Applications to data for Galactic globular clusters and nearby
dE galaxies will be presented in future papers. These are only two
examples of a large range of dynamical problems in astronomy to which
a discrete \sch code like ours can be applied, so we expect this new
tool will contribute to the better understanding of stellar systems in
general.

%\begin{acknowledgements}
%Thanks to  ....
%\end{acknowledgements}
%
%\section{Acknowledgements}
%\label{sec:gracias}

\acknowledgements

We are happy to thank Marla Geha and Raja Guhathakurta for their
continued interest in the present work and its extension to the study
of actual galaxies using their unique data on dwarf ellipticals. We
also thank Glenn van de Ven for very useful discussions, his interest
in the progress of this project and, last but not least, for his
invaluable help with IDL routines. This paper also benefited by
comments from Davor Krajnovic, Aaron Romanowsky, and David
Merritt. Thanks also to George Meylan for his help with the writing of
the HST Theory proposal specified below, and to the anonymous referee,
whose comments and suggestions improved the presentation of the
paper. This work was carried out as part of HST Theory Project \#9952
and was supported by NASA through a grant from STScI, which is
operated by AURA, Inc., under NASA contract NAS 5-26555.



%\appendix

%\section{Appendix material}

\clearpage

%\begin{landscape}
%\rotate
\begin{deluxetable}{ccrlcrlcrlcrlcrl}
\tablewidth{0pc}
%\tabletypesize{\tiny}
\tabletypesize{\scriptsize}
\tablecaption{Comparison between input and fitted orbital mass weights}
\tablehead{
\multicolumn{1}{c}{dataset} &
\multicolumn{1}{c}{} &
\multicolumn{2}{c}{no projection} &
\multicolumn{1}{c}{} &
\multicolumn{2}{c}{$I_3$} &
\multicolumn{1}{c}{} &
\multicolumn{2}{c}{$L_z,I_3$} &
\multicolumn{1}{c}{} &
\multicolumn{2}{c}{$E,I_3$} &
\multicolumn{1}{c}{} &
\multicolumn{2}{c}{$E,L_z$} \\
\multicolumn{1}{c}{} &
\multicolumn{1}{c}{} &
\multicolumn{1}{c}{RMS} &
\multicolumn{1}{c}{$\mid{\rm med}\mid$} &
\multicolumn{1}{c}{} &
\multicolumn{1}{c}{RMS} &
\multicolumn{1}{c}{$\mid{\rm med}\mid$} &
\multicolumn{1}{c}{} &
\multicolumn{1}{c}{RMS} &
\multicolumn{1}{c}{$\mid{\rm med}\mid$} &
\multicolumn{1}{c}{} &
\multicolumn{1}{c}{RMS} &
\multicolumn{1}{c}{$\mid{\rm med}\mid$} &
\multicolumn{1}{c}{} &
\multicolumn{1}{c}{RMS} &
\multicolumn{1}{c}{$\mid{\rm med}\mid$}
}
\startdata
55ns & & 4.5138 & 0.4531 & & 0.2207 & 0.1404 & & 0.0908 & 0.0359 & & 0.1080 & 0.0719 & & 0.2634 & 0.1591 \\
55is & & 6.6525 & 0.5630 & & 0.3677 & 0.1915 & & 0.0939 & 0.0309 & & 0.1912 & 0.1598 & & 0.2460 & 0.1855 \\
55ms & & 3.1524 & 0.4208 & & 0.2123 & 0.1207 & & 0.0884 & 0.0334 & & 0.1845 & 0.0903 & & 0.2505 & 0.1122 \\
90ns & & 3.7560 & 0.5939 & & 0.2683 & 0.1633 & & 0.1419 & 0.0425 & & 0.2202 & 0.1365 & & 0.2361 & 0.2491 \\
90is & & 2.7956 & 0.6410 & & 0.6253 & 0.1629 & & 0.1366 & 0.0347 & & 0.4826 & 0.1178 & & 0.2566 & 0.2364 \\
90ms & & 1.4893 & 0.4927 & & 0.2298 & 0.1714 & & 0.1216 & 0.0384 & & 0.1456 & 0.1149 & & 0.2953 & 0.2333 \\
%\\
%55ns & & 8.7257 & 0.4920 & & 0.3179 & 0.1459 & & 0.0304 & 0.0173 & & 0.1512 & 0.1022 & & 0.5774 & 0.1500 \\
%55is & & 9.3958 & 0.6237 & & 0.5400 & 0.2400 & & 0.0316 & 0.0171 & & 0.2450 & 0.2281 & & 0.7361 & 0.1604 \\
%55ms & & 4.5211 & 0.4539 & & 0.2914 & 0.1731 & & 0.0305 & 0.0181 & & 0.2278 & 0.1553 & & 0.6633 & 0.1909 \\
%90ns & & 4.1376 & 0.5866 & & 0.4036 & 0.1687 & & 0.0481 & 0.0375 & & 0.2173 & 0.1300 & & 0.7313 & 0.2064 \\
%90is & & 5.8993 & 0.7236 & & 2.5684 & 0.2071 & & 0.0452 & 0.0324 & & 0.7889 & 0.1608 & & 0.7603 & 0.1658 \\
%90ms & & 1.7923 & 0.5176 & & 0.3792 & 0.1960 & & 0.0290 & 0.0251 & & 0.1682 & 0.0896 & & 0.8509 & 0.1511 \\
\enddata

\tablecomments{The tabulated numbers are the root mean square and
median absolute deviation of the quantity $(\zeta_{\rm fit}-\zeta_{\rm
in})/\zeta_{\rm in}$, i.e., the difference between fit and input mass
weights normalized by the input mass weights, for \sch models based on
our small ($20\times14\times7$) orbit library. The statistics are
always computed inside the energy range constrained by the data (see
Fig.\ref{fig:1Dplots}), and are shown for the full cubes of mass
weights (columns labeled ``no projection'') and for various
projections of these cubes. The projected distributions are obtained
by integrating over one or two of the integrals of motion (i.e., by
collapsing the 3-D cubes in one or two dimensions), and appear under
the columns labeled by the integral(s) of motion over which the
integration has been done. }

\end{deluxetable}
%\end{landscape}

\clearpage

%% the file, uncomment the \plotone, \plottwo, and
%% If you need a layout that cannot be achieved with \plotone or
%% \plottwo, you can invoke the graphicx package directly with the
%% \includegraphics command or use \plotfiddle. For more information,
%% please see the tutorial on "Using Electronic Art with AASTeX" in the
%% documentation section at the AASTeX Web site,
%% http://www.journals.uchicago.edu/AAS/AASTeX.
%%
%% The examples below also include sample markup for submission of
%% supplemental electronic materials. As always, be sure to check
%% the instructions to authors for the journal you are submitting to
%% for specific submissions guidelines as they vary from

%% This example uses \plotone to include an EPS file scaled to
%% 80% of its natural size with \epsscale. Its caption
%% has been written to indicate that additional figure parts will be
%% available in the electronic journal.

%%\begin{figure}
%%\vspace*{-1cm}
%%%{\hspace{-15cm}
%%%\plotfiddle{positions_55is_1E5.ps}{15cm}{0}{50}{50}{-25}{0}}
%%\vspace*{-1cm}
%%\caption{\label{fig:data55is} Phase-space projections of the 55is
%%dataset. }\end{figure}

\clearpage

\clearpage

\clearpage

\clearpage

\clearpage

\clearpage

\clearpage

\clearpage

\clearpage 

\clearpage

\clearpage

\pagestyle{empty}

\clearpage

\end{document}